\documentclass[11pt,a4paper]{article}
\pdfoutput=1

\usepackage{jheppub}
                     
\usepackage[mathscr]{euscript}
\usepackage[textsize=scriptsize,textwidth=2.5cm]{todonotes}

\usepackage{amssymb,amsmath,bm,amsfonts}
\usepackage{graphicx}
\usepackage{longtable}
\usepackage{color,xcolor}
\usepackage{subfig}
\usepackage{comment}
\usepackage{float}
\usepackage[normalem]{ulem}
\usepackage{array}

\usepackage{relsize}
\usepackage{graphicx}

\usepackage{etoolbox}
\appto\appendix{\addtocontents{toc}{\protect\setcounter{tocdepth}{1}}}

\appto\listoffigures{\addtocontents{lof}{\protect\setcounter{tocdepth}{1}}}
\appto\listoftables{\addtocontents{lot}{\protect\setcounter{tocdepth}{1}}}

\numberwithin{equation}{section}
\allowdisplaybreaks

\makeatletter
\def\@fpheader{\phantom{Prepared for submission to JHEP}}
\makeatother

\newcommand{\bea}{{\begin{eqnarray}}}
\newcommand{\eea}{{\end{eqnarray}}}
\newcommand{\mA}{{\cal A}}
\newcommand{\mB}{{\cal B}}
\newcommand{\mC}{{\cal C}}
\newcommand{\mH}{{\cal H}}
\newcommand{\mM}{{\cal M}}
\newcommand{\zc}{{z}}

\newcommand{\bra}[1]{\ensuremath{\left\langle#1\right|}}
\newcommand{\ket}[1]{\ensuremath{\left|#1\right\rangle}}

\newcommand{\be}{\begin{equation}}
\newcommand{\ee}{\end{equation}}
\newcommand{\bpm}{\begin{pmatrix}}
\newcommand{\epm}{\end{pmatrix}}

\newcommand{\EV}[1]{\langle #1 \rangle}
\newcommand{\beqn}{\begin{eqnarray}}
\newcommand{\eeqn}{\end{eqnarray}}

\newcommand{\p}{\partial}

\newcommand{\ba}{\begin{aligned}}
\newcommand{\ea}{\end{aligned}}
\newcommand{\bi}{\begin{enumerate}}
\newcommand{\ei}{\end{enumerate}}

\def\Vev#1{\left\langle#1\right\rangle}
\newcommand{\eq}[1]{\begin{align}#1\end{align}}
\newcommand{\eqsp}[1]{\begin{equation}\begin{split}#1\end{split}\end{equation}}

\DeclareMathOperator{\tr}{tr}

\title{Swing surfaces and holographic entanglement beyond AdS/CFT}

\author[a]{Luis Apolo,} 
\author[b]{Hongliang Jiang,} 
\author[a,c]{Wei Song} 
\author[a]{and Yuan Zhong} 

\affiliation[a]{Yau Mathematical Sciences Center, Tsinghua University, Beijing 100084, China}
 
\affiliation[b]{Albert Einstein Center for Fundamental Physics, Institute for Theoretical Physics, University of Bern, Sidlerstrasse 5, 3012 Bern, Switzerland}

\affiliation[c]{Institute for Advanced Study, 1 Einstein Drive, Princeton, NJ 08540, USA}
  \emailAdd{apolo@mail.tsinghua.edu.cn, jiang@itp.unibe.ch, wsong2014@mail.tsinghua.edu.cn, zhongy17@mails.tsinghua.edu.cn} 
 
\abstract{We propose a holographic entanglement entropy prescription for general states and regions in two models of holography beyond AdS/CFT known as flat$_3$/BMSFT and (W)AdS$_3$/WCFT. Flat$_3$/BMSFT is a candidate of holography for asymptotically flat three-dimensional spacetimes, while (W)AdS$_3$/WCFT is relevant in the study of black holes in the real world. In particular, the boundary theories are examples of quantum field theories that feature an infinite dimensional symmetry group but break Lorentz invariance. Our holographic entanglement entropy proposal is given by the area of a {\it swing surface} that consists of {\it ropes}, which are null geodesics emanating from the entangling surface at the boundary, and a {\it bench}, which is a spacelike geodesic connecting the ropes. The proposal is supported by an extension of the Lewkowycz-Maldacena argument, reproduces previous results based on the Rindler method, and satisfies the first law of entanglement entropy.}

\begin{document}
\maketitle
\flushbottom

\section{Introduction}

Understanding the emergence of spacetime is a profound question in the holographic description of quantum gravity. The seminal work of Ryu-Takayanagi (RT)~\cite{Ryu:2006bv} and Hubeny-Rangamani-Takayanagi (HRT)~\cite{Hubeny:2007xt} showed that entanglement entropy in conformal field theories is dual to the area of extremal surfaces in asymptotically AdS spacetimes in Einstein gravity. This deep connection between spacetime geometry and quantum entanglement has motivated the idea that entanglement builds geometry~\cite{Swingle:2009bg,VanRaamsdonk:2009ar,VanRaamsdonk:2010pw}, and has led to fruitful connections between quantum information, condensed matter physics, and black holes.

The RT/HRT proposal has been firmly established~\cite{Casini:2011kv,Lewkowycz:2013nqa,Dong:2016hjy} within the framework of the AdS/CFT correspondence and it is natural to ask whether this connection between  entanglement and geometry is more general and can be extended beyond AdS/CFT. Efforts toward this direction include studies on warped AdS$_3$~\cite{ Anninos:2013nja,Castro:2015csg, Song:2016pwx,Song:2016gtd,Azeyanagi:2018har,Wen:2018mev,Apolo:2018oqv,Chen:2019xpb,Gao:2019vcc},  flat~\cite{Bagchi:2014iea,Basu:2015evh,Hosseini:2015uba,Jiang:2017ecm,Hijano:2017eii,Godet:2019wje,Fareghbal:2019czx}, Lifshitz~\cite{Gentle:2017ywk,Gentle:2015cfp}, and de Sitter spacetimes~\cite{Sanches:2016sxy,Dong:2018cuv,Lewkowycz:2019xse}, among others.  In the literature, there are several approaches to study holographic entanglement entropy in these backgrounds. One approach consists of studying properties of the RT/HRT surfaces with the assumption that the RT/HRT proposal extends directly to non-AdS spacetimes. The main advantage of this approach is having a concrete geometric picture in the bulk, although the minimal area prescription is not a priori guaranteed to reproduce the entanglement entropy of the dual field theory. Alternatively, the approach taken in~\cite{Castro:2015csg,Song:2016gtd,Jiang:2017ecm} consists of {\it deriving} the bulk dual of the boundary entanglement entropy by generalizing the Rindler method~\cite{Casini:2011kv} to the conjectured holographic description of non-AdS spacetimes.\footnote{In this paper, by non-AdS spacetimes we refer to backgrounds that do not satisfy the standard asymptotically AdS (Dirichlet) boundary conditions.} The advantage of this approach is that the bulk computation of entanglement entropy is automatically consistent with the holographic dictionary, while its disadvantage is that the Rindler method is only applicable for special entangling regions and special states. In this paper, we propose a general geometric prescription for entanglement entropy in a class of non-AdS spacetimes and show that it is consistent with the generalized gravitational entropy \'{a} la~\cite{ Lewkowycz:2013nqa,Dong:2016hjy}.
  
\begin{figure}[ht!] 
   \centering
    \includegraphics[scale=0.6875]{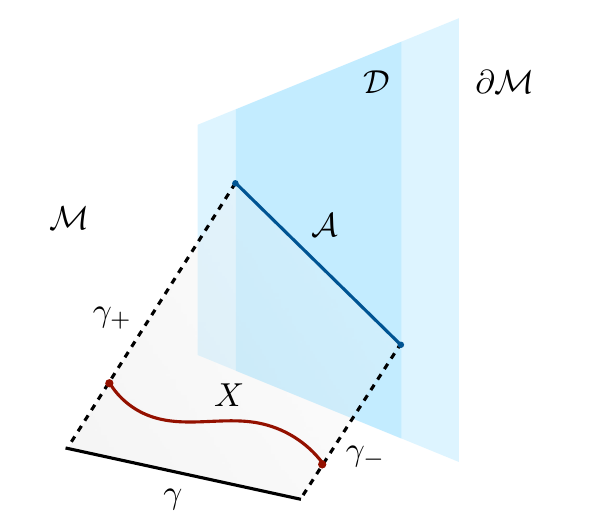} 
    \caption{Sketch of two surfaces $X_{\mA}$ in the bulk $({\cal M})$ that are homologous to the interval $\mA$ at the boundary $(\partial {\cal M})$. These surfaces consist of the union of either $X$ or $\gamma$ and the collection of null geodesics $\gamma_{b\partial} = \gamma_{+} \cup \gamma_{-}$ connecting the bulk to the boundary. The $X$ and $\gamma$ surfaces lie between $\gamma_{+}$ and $\gamma_{-}$ but only $\gamma$ is an extremal surface.}
\label{s2:swingsurface} 
\end{figure}

Let us consider a holographic duality between a $d$-dimensional quantum field theory and a  gravitational theory in $d+1$ dimensions that reduces to Einstein gravity in the semiclassical limit. We assume that consistent boundary conditions exist such that the symmetries and partition functions of the bulk and boundary theories match. In addition, we assume that the modular Hamiltonian of the vacuum state can be geometrically realized as the generator of a diffeomorphism. We then propose that the entanglement entropy of a co-dimension one subregion $\mA$ in the field theory at the boundary can be obtained from the bulk theory of gravity by 
\eq{
S_{\mA} = \min  \underset {X_{\cal A}\sim \cal A }{\textrm{ext}}  \frac{\textrm{Area}(X_{\cal A} )}{4G},\qquad X_{\cal A} =X \cup_{p\in\p\mA}\gamma_{(p)},   \label{proposalintro}
}
where $X_\mA$ is a surface homologous to $\cal A$ that consists of a spacelike surface $X$ and a collection of null geodesics $\gamma_{(p)}$ (see Fig.~\ref{s2:swingsurface}).
 The null geodesic $\gamma_{(p)}$ emanates from a point $p$ at the boundary $\p\mA$ of the entangling surface $\mA$, and its tangent vector reduces to the approximate modular flow generator near $p$ after an appropriate normalization; while the surface $X$ is an arbitrary co-dimension two spacelike surface connecting the ropes. The entanglement entropy is determined by first extremizing the area of all possible surfaces $X$ and then choosing the minimal area.  The resulting configuration is referred to as a {\it swing surface} which is denoted by $\gamma_{\cal A} = \gamma \cup_{p\in\p\mA}\gamma_{(p)}$ where $\gamma$ is a spacelike {\it bench} connecting the null {\it ropes} $\gamma_{(p)}$.

The proposal \eqref{proposalintro} is compatible with the RT/HRT prescription in the AdS/CFT correspondence. In this case, the ropes $\gamma_{(p)}$ shrink towards the boundary and the swing surface reduces to the well-known RT/HRT surface. In contrast, in the examples of non-AdS holography considered in this paper, the crucial difference between \eqref{proposalintro} and the RT/HRT proposal  is that the spacelike surface $X$ is connected to the boundary interval $\mA$ by the null ropes $\gamma_{(p)}$. The ropes are necessary to satisfy the homology constraint and to consistently extend the replica trick from the boundary to the bulk. This extension of the replica trick leads to a generalization of the Lewkowycz-Maldacena argument~\cite{Lewkowycz:2013nqa} and it allows us to show that the generalized gravitational entropy localizes on the swing surface, thereby providing a derivation of our proposal. Using the swing surface proposal~\eqref{proposalintro}, we also extend the concept of relative entropy to non-AdS holography and verify the first law of entanglement entropy.

As explicit examples, we study the proposal~\eqref{proposalintro} in two models of non-AdS holography dubbed flat$_3$/BMSFT and (W)AdS$_3$/WCFT. In the first model, flat$_3$ refers to three-dimensional asymptotically flat spacetimes whose asymptotic symmetry group is the three-dimensional version of the Bondi-van der Burg-Metzner-Sachs (BMS) group~\cite{Bondi:1962px,Sachs:1962wk,Sachs:1962zza}. Hence, the phase space of gravity in these backgrounds is organized by the same symmetry group as that of a two-dimensional BMS-invariant field theory (BMSFT)~\cite{Barnich:2006av,Barnich:2010eb,Bagchi:2010eg,Bagchi:2012cy}. In the second model, WCFT refers to a warped conformal field theory --- a two-dimensional quantum field theory characterized by a Virasoro-Kac-Moody algebra~\cite{Hofman:2011zj,Detournay:2012pc} --- whose bulk description is gravity on either asymptotically AdS$_3$ spacetimes with Dirichlet-Neumann boundary conditions~\cite{Compere:2013bya} or asymptotically warped AdS$_3$ backgrounds~\cite{Anninos:2008fx,Compere:2009zj}.

It has been noticed in~\cite{Song:2016gtd,Jiang:2017ecm} that swing surfaces appear as the geometric description of holographic entanglement entropy in flat$_3$/BMSFT and (W)AdS$_3$/WCFT for single intervals on zero-mode backgrounds. In a companion paper \cite{Apolo:2020qjm}, we work out the modular Hamiltonian above zero-mode backgrounds from the field theory side and match it to a gravitational charge in the bulk. These studies are based on a generalization of the Rindler method \cite{Casini:2011kv, Castro:2015csg} or, equivalently, on the existence of a local modular Hamiltonian. In this paper we propose that swing surfaces work for more general intervals and more general backgrounds, and provide a Lewkowycz-Maldacena type argument in support of this proposal. We will illustrate how to apply the general prescription put forward in Section \ref{se:proposal} to both flat$_3$/BMSFT and (W)AdS$_3$/WCFT without resorting to the Rindler method or the existence of a global  modular flow generator. Using this new approach, we reproduce the previous results of~\cite{Song:2016gtd,Jiang:2017ecm} on zero mode backgrounds and obtain new results on perturbative states above these backgrounds. 

The paper is organized as follows. In Section~\ref{se:generalproposal} we propose a general prescription for the computation of holographic entanglement entropy in terms of swing surfaces. Therein we extend relative entropy to models of non-AdS holography and check that the first law of entanglement entropy holds. In addition, we comment on the generalization of our results to multiple intervals, the entanglement wedge associated with the swing surface, and deformations of the swing surface. In Sections~\ref{se:flat} and~\ref{se:warped} we test the proposal in two models of non-AdS holography, namely the flat$_3$/BMSFT and (W)AdS$_3$/WCFT correspondences. In particular, we reproduce previous results on holographic entanglement entropy for arbitrary zero-mode backgrounds in both of these models. Furthermore, in Section~\ref{se:flat} we extend these results to linearized perturbations of asymptotically flat three-dimensional spacetimes and discuss strong subadditivity in flat$_3$/BMSFT. We collect several additional results in Appendices~\ref{app:extremal} --~\ref{app:geodesicsads}. In Appendix~\ref{app:extremal} we prove that in cases where the bulk modular flow is an exact Killing vector, the bench of the swing surface is an extremal surface; in Appendix~\ref{app:nullgeodesics} we describe general results on the parametrization of null geodesics; in Appendix~\ref{app:aproxmodflowbulk} we extend the approximate modular flow generator into the bulk; in Appendix~\ref{app:zeroMode2} we give additional details on the swing surface of three-dimensional flat spacetimes; and in Appendix~\ref{app:geodesicsads} we describe null geodesics in locally AdS$_3$ spacetimes.


\section{A general proposal for holographic entanglement entropy} \label{se:generalproposal}
In this section we propose a geometric prescription for entanglement entropy in models of holography beyond AdS/CFT. These models consist of a $(d+1)$-dimensional theory of gravity in the bulk and a $d$-dimensional field theory at the boundary satisfying the following conditions:
\begin{enumerate}\label{assumption}

\item[(1)] The bulk theory of gravity admits a semiclassical description in terms of Einstein gravity. 

\item[(2)]  The field theory is invariant under a symmetry group $\sf G$ and the vacuum state is invariant under a subgroup of $\sf G$ whose generators are denoted by $h_i$. 

\item[(3)] Consistent boundary conditions exist such that the asymptotic symmetry group in the bulk agrees with $\sf G$ at the boundary. 

\item[(4)] The bulk theory admits a solution with Killing vectors $H_i$ that  correspond to the generators $h_i$ at the boundary. This bulk geometry is identified with the vacuum state in the dual field theory.

\item[(5)] The partition function in the bulk theory of gravity agrees with the partition function of the field theory at the boundary.

\end{enumerate}
In analogy with the AdS/CFT correspondence, we expect that additional conditions are necessary to guarantee a consistent holographic correspondence, see e.g.~\cite{ElShowk:2011ag} for a review. In addition, we require that
\begin{enumerate}
\item[(6)] A local modular Hamiltonian can be written down for ball-shaped regions (single intervals in $d=2$) on the vacuum, as discussed in more detail in Section~\ref{se:rindler}.
\end{enumerate}

All of the properties above are satisfied in the AdS/CFT correspondence, as well as in models of non-AdS holography including flat$_3$/BMSFT and (W)AdS$_3$/WCFT. In particular, condition (6) provides a geometric description of entanglement entropy for special regions on the vacuum that corresponds to the RT/HRT surface in AdS/CFT~\cite{Casini:2011kv}, and a swing surface in both flat$_3$/BMSFT and (W)AdS$_3$/WCFT~\cite{Song:2016gtd,Jiang:2017ecm,Apolo:2020qjm}. In the following, we will first describe the swing surface on the vacuum state and then extend it to general states and shapes of the entangling surface.  We also explore the consequences of our proposal to the relative entropy of general states, the entanglement wedge, and the entanglement entropy of multiple intervals.


\subsection{Modular flow and swing surfaces on the vacuum }\label{se:rindler}
In this section we review previous results on holographic entanglement entropy on the vacuum in models of holography satisfying the properties listed above~\cite{Song:2016gtd,Jiang:2017ecm,Apolo:2020qjm}. In particular, we will show that the geometric picture of entanglement entropy in the bulk is a swing surface which consists of a co-dimension two spacelike surface that hangs from the subregion $\mA$ at the boundary by a collection of null geodesics. 

Let us consider a co-dimension one subregion $\mA$ on the $d$-dimensional boundary $\p \cal M$ of the bulk $(d+1)$-dimensional spacetime $\cal M$. The entanglement entropy of subregion $\mA$ on a state $\ket{\psi}$ is given by the von Neumann entropy
\eq{
S_{\mA} = - \tr (\rho_\mA \log \rho_\mA),
}
where $\rho_\mA = \tr_{\bar{\cal A}} \ket{\psi}\bra{\psi}$ is the reduced density matrix of subregion $\mA$ obtained by tracing over the degrees of freedom on the complement $\bar{\cal A}$. The entanglement entropy can be understood as a thermal entropy with respect to the modular Hamiltonian ${\cal H}_{mod}$ such that 
  \eq{
	\rho_\mA = e^{-\mH_{mod}}. \label{s2:modHam}
  }
The modular Hamiltonian generates a modular flow in the causal development of subregion $\mA$ and is generically nonlocal. In relativistic QFTs,  examples where a local modular Hamiltonian can be written down are related to the Rindler wedge of the vacuum~\cite{Bisognano:1975ih,Bisognano:1976za}, which has led to fruitful results in CFTs~\cite{Casini:2011kv,Cardy:2016fqc,Casini:2017roe}. This approach has also been extended to BMSFTs and WCFTs in~\cite{Song:2016gtd,Jiang:2017ecm,Apolo:2020qjm}. 
 
In the aforementioned examples, the modular Hamiltonian generates a flow that is geometrically realized and corresponds to a symmetry of the underlying quantum field theory. In particular, the modular flow generator can be written as
\eq{
\zeta =\sum_{i} a_i h_i \equiv 2\pi \p_\tau, \label{s2:modflowboundary}
}
where $h_i$ are the vacuum symmetry generators of the quantum field theory, $a_i$ are parameters that depend on subregion $\mA$, and we have introduced the Rindler time coordinate $\tau$. As shown for two-dimensional CFTs in~\cite{Czech:2019vih} and both BMSFTs and WCFTs in~\cite{Apolo:2020qjm}, the coefficients $a_i$ can be uniquely determined by requiring that ($i$) for any real parameter $s$, the flow $e^{s \zeta}$ maps any point in the causal domain $\cal D$ of $\mA$ to another point in $\cal D$ and leaves the boundary $\p\cal D$ of the causal domain invariant; and ($ii$) the flow $e^{i \zeta}$ maps any point in $\cal D$ back to itself and hence generates the thermal identification
\eq{
 \tau \sim \tau + 2 \pi i. \label{s2:temperature}
}
The modular flow generator~\eqref{s2:modflowboundary} can also be interpreted as generating translations along a local Rindler time coordinate $\tau$. The origin of this interpretation can be traced back to the existence of a symmetry transformation that maps subregion $\mA$ in $\p \cal M$ to a noncompact generalized Rindler spacetime $\p \tilde {\cal M}$, as described in more detail in~\cite{Castro:2015csg,Song:2016gtd,Jiang:2017ecm}.

According to our assumptions, the vacuum state in the field theory is dual to a spacetime in the bulk that is invariant under the same set of symmetries. In particular, the vacuum symmetry generators $h_i$ at the boundary correspond to the set of Killing vectors $H_i$ in the bulk satisfying $H_i |_{\p\mM} = h_i$, or more explicitly, $H_i^a |_{r={1/\varepsilon}}=h_i^a$ where $r = 1/\varepsilon$ is a cutoff surface along the bulk radial coordinate and $a$ labels the coordinates at the boundary. Thus, the modular flow generator at the boundary~\eqref{s2:modflowboundary} can be expressed as a linear combination of Killing vectors in the bulk such that
  \eq{
  \xi =\sum_{i} a_i H_i\equiv 2\pi \p_{\tau}, \qquad \tau\sim\tau + 2 \pi i, \label{s2:modflowbulkexact}
  }
where the coefficients $a_i$ are the same coefficients featured in \eqref{s2:modflowboundary}. Note that the Rindler time function $\tau$ defined in \eqref{s2:modflowbulkexact} is a natural extension of the boundary Rindler time \eqref{s2:modflowboundary} into the bulk, and hence it satisfies the same periodicity condition~\eqref{s2:temperature}. By abuse of notation, we use the same variable $\tau$ to denote the Rindler time function in both the boundary and the bulk. 

The thermal identification in \eqref{s2:modflowbulkexact} implies that the bulk modular flow generator $\xi$ features a bifurcating Killing horizon with surface gravity $2\pi$.  Let us denote the bifurcating surface by $\gamma_\xi$ and the outward half (towards the boundary) of the future/past Killing horizon by $N_{\pm}$. Then, $N_\pm$ corresponds to the future/past light-sheet of $\gamma_\xi$ and the modular flow generator~\eqref{s2:modflowbulkexact} satisfies 
  \eq{
 & \xi \big|_{\gamma_\xi} = 0 , \qquad \nabla^\mu \xi^\nu \big|_{\gamma_\xi} = 2\pi n^{\mu\nu}, \label{s2:xip1}\\
&\xi^{\nu} \nabla_{\nu} \xi^{\mu} \big|_{N_{\pm}} = \pm 2\pi \xi^{\mu},   \label{s2:xip2}
}
where $n^{\mu\nu}$ is the binormal unit vector of $\gamma_\xi$. From Frobenius' theorem, as well as the fact that $\xi$ is normal to the light-sheets $N_{\pm}$, it follows that
\eq{
\xi_{[\mu} \nabla_\nu \xi_{\lambda]}\big|_{N_\pm}=0.
}
For more details on the bulk modular flow generator in non-AdS holography  see~\cite{Apolo:2020qjm}.

In models of non-AdS holography like flat$_3$/BMSFT and (W)AdS$_3$/WCFT, the fixed points of the boundary modular flow generated by $\zeta$ are not the fixed points of the bulk modular flow generated by $\xi$. This means that the bifurcating surface $\gamma_\xi$ is not attached to the interval $\mA$ at the boundary. Instead, the homology condition is satisfied by a generalization of the RT surface $\gamma_{\mA}$ that is defined by 
  \eq{
   \gamma_\mA = \gamma \cup \gamma_{b\p},   \qquad   \gamma_{b\p}\equiv \cup_{p\in\p\mA}\gamma_{(p)},  \label{swingsurface}
  }
where $\p\mA$ is the boundary of the subregion $\mA$,  $\gamma_{(p)}$ is the trajectory of each point $p\in\p\mA$ generated by the bulk modular flow, and $\gamma$ is a subregion of the bifurcating surface $\gamma_\xi$ that is bounded by the intersection of $\gamma_\xi$ with $\gamma_{b\p}$. Since $\p\mA$ belong to the null surface $N_+\cup N_-$, the latter of which is invariant under the bulk modular flow, we learn that the entire trajectory $\gamma_{(p)}$ lies on $N_+\cup N_-$. Furthermore, since the tangent vector of $\gamma_{(p)}$ is parallel to $\xi$, which is by definition null on $N_\pm$, we conclude that $\gamma_{(p)}$ is a null geodesic. 

We refer to surfaces like $\gamma_{\cal A}$ that consist of a spacelike {\it bench} $\gamma$ and a collection of null {\it ropes} $\gamma_{b\p}$ as {\it swing surfaces}. In particular, for single intervals in flat$_3$/BMSFT and (W)AdS$_3$/WCFT we have $\gamma_{b\p}=\gamma_+\cup\gamma_-$ where $\gamma_\pm \subset N_{\pm}$ are null geodesics connecting the endpoints of the interval $\mA$ at the boundary to the endpoints of the spacelike surface $\gamma$ in the bulk. In these examples, the shape of $\gamma_{\cal A}$ indeed looks like a swing hanging from the boundary at $\p \cal A$. As shown in Appendix~\ref{app:extremal}, the bench $\gamma$ extremizes the distance between the ropes $\gamma_+$ and $\gamma_-$. Consequently, the swing surface $\gamma_\mA$ plays the role of the RT/HRT surface in these models of non-AdS holography. The entanglement entropy of the boundary interval $\mA$ is then given by the area of the swing surface in Planck units~\cite{Song:2016gtd,Jiang:2017ecm}
\eq{
S_\mA ={ \mathrm{Area}(\gamma_{\cal A} )\over 4G}.
}

To summarize, for states where the bulk modular flow generator is an exact Killing vector, the ropes of the swing surface are null geodesics generated by the modular flow while the bench of the swing surface is the set of fixed points of the modular flow generator that extremizes the distance between the ropes.


\subsection{Approximate modular flow for general states} \label{se:localmodflow}
For general states and/or shapes of the entangling surface, the bulk modular flow is no longer generated by an exact Killing vector of the background geometry. In this section we find the vector field that generates the modular flow near the swing surface by generalizing the approach of \cite{Lashkari:2016idm} to models of non-AdS holography satisfying the conditions described at the beginning of Section~\ref{se:generalproposal}.

\subsubsection*{The approximate modular flow generator in the boundary}

Once we have the global modular flow generator $\zeta$ for the vacuum state, we can generalize it into a local version that is valid for more general states and shapes of the entangling surface $\mA$. Note that the boundary $\p\mA$ of $\mA$ is a co-dimension two surface in ${\p\mM}$. At each point $p\in\p \mA$ there is a two-dimensional plane transverse to $\p \mA$ within which it is always possible to find a local Rindler coordinate system in a small neighborhood near $p$.\footnote{For $d=2$ the normal plane refers to the entire manifold at the boundary, namely $\p\mM$.} In other words, it is always possible to find an approximate modular flow generator 
\eq{
\zeta^{(p)} = \sum_i a^{(p)}_i h_i  = 2\pi \p_{\tau_{(p)}},\label{s2:aproxmodflow}
}
where $h_i$ are the symmetry generators in the transverse plane near $p$ and $\tau_{(p)}$ parametrizes time in the local Rindler coordinates. We assume that all the $\tau_{(p)}$ can be smoothly sewn together close to $\p\cal A$ so we will drop the subscript $(p)$ henceforth. When $d = 2$, $\p \mA$ is a set of isolated points. In this case, for each endpoint $p\in\p \mA$, the approximate modular flow generator and the associated Rindler coordinate transformation can be obtained from the corresponding expressions for single intervals on the vacuum by sending the other endpoint to infinity. In higher dimensions, $\p\mA$ is a co-dimension two surface in $\p \cal M$. We assume that appropriate gluing conditions exist such that the approximate modular flow generator can be smoothly defined on a small neighborhood of $\p\mA$.

The approximate modular flow generator $\zeta^{(p)}$ and the local Rindler time $\tau$ defined near $\p \mA$ can be used to define measures of entanglement using the replica trick. In analogy with the discussion in CFTs, we expect that the replica trick can be performed on the field theory at the boundary by making $n$ copies of $\p \mM$ and gluing them cyclically along $\cal A$. The replica symmetry can then be implemented by imposing periodic boundary conditions around each point $p\in\p \cal A$ such that
\eq{
\tau\sim \tau+2\pi i n, \qquad \phi_{\p\mM} (\tau+2 \pi  i n) = \phi_{\p\mM}(\tau),\label{periodbdy}
}
where $\phi_{\p\mM}$ collectively denote the fields in the boundary field theory.\footnote{This picture is valid for states symmetric under time reversal. For more general states, we need to sew different Rindler wedges together and use the Schwinger-Keldysh formalism, in analogy with the derivation of the HRT proposal~\cite{Dong:2016hjy}.}  
Consequently, the R\'{e}nyi and entanglement entropies can be defined in a similar fashion to CFTs, namely,  
 \eq{
{S}_n&= {1\over 1-n}Z_n,\qquad Z_n \equiv \tr \rho_{\cal A}^n=\tr e^{-n \mathcal H_{mod}}, \\
S_{\cal A}&= \lim_{n\to1} S_n=\lim_{n\to1}n \p_n\big( n\log Z_1 -\log Z_n \big). \label{EEbn}
  }


\subsubsection*{The approximate modular flow generator in the bulk}

Let us now extend the approximate modular flow generator $\zeta^{(p)}$ into the bulk. The first step is to extend the $d$-dimensional vector field $\zeta^{(p)}$ into a $(d+1)$-dimensional vector field $\xi^{(p)}_\infty$ near the asymptotic boundary. The vector field $\xi^{(p)}_\infty$ is given by
  \eq{
  \xi^{(p)}_\infty=\sum_i a^{(p)}_i H_i, \label{s2:aproxmodflowbulk}
  }
   where $H_i$ are the approximate Killing vectors near $p \in\p\mA$ satisfying $H_i|_{{\p \mM}}=h_i$ and the coefficients $a_i^{(p)}$ are the same coefficients featured in \eqref{s2:aproxmodflow}. It follows that $\xi^{(p)}_\infty|_{\p\mM}=\zeta^{(p)}$.
   
The next step is to extend the $(d+1)$-dimensional vector field $\xi^{(p)}_\infty$ defined near the asymptotic boundary into the bulk.  As described in the previous section, when the bulk modular flow generator is an exact Killing vector, the modular flow moves each point $p\in\p\mA$ into the bulk along a null geodesic. When the bulk modular flow is no longer generated by a Killing vector, we still expect it to move a point $p\in \p\mA$ along a null geodesic until it reaches a fixed point in the bulk, if the latter exists. A starting point $p$ and a null vector $\xi^{(p)}_\infty$ uniquely determine a null geodesic $\gamma_{(p)}$ towards the bulk.  This property enables us to extend $\xi^{(p)}_\infty$ defined near the asymptotic boundary to a null vector field $\xi^{(p)}$ defined along $\gamma_{(p)}$ satisfying
  \eq{
  \xi^{(p)\mu} \xi^{(p)}_\mu=0, \qquad \xi^{(p)}\big|_{\p\mM}=\zeta^{(p)}. \label{s2:xipbc}
}
In Appendix~\ref{app:nullgeodesics}, we show that in a Lorentzian manifold it is always possible to parametrize the null geodesic $\gamma_{(p)}$ in terms of coordinates $x^{\mu}(\lambda)$ and $x^{\mu}(\tau)$ --- where $\lambda$ is an affine parameter and $\tau$ the local Rindler time --- such that the tangent vector $\xi^{(p)\mu}= 2\pi {d x^{\mu} / d \tau} = 2\pi (d\lambda/d\tau) (d x^\mu /d\lambda)$ satisfies
\eq{
\quad \xi^{(p)\mu}\nabla_{\mu}\xi^{(p)\nu}=  \pm 2\pi \xi^{(p)\nu}, \label{s2:nmbk}
}
where the $\pm$ sign indicates that the vector is outward or inward directed. The parametrization above shows that the parameter $\tau$ defined via $2\pi\p_\tau=\xi^{(p)}$ has identification $\tau\sim\tau+2\pi i$. Hence, $\xi^{(p)}$ naturally extends the local Rindler time at the boundary into the bulk along the null geodesics. Furthermore, note that the null vector field $\xi^{(p)}$ defined in this way has a fixed point $\p \gamma_{(p)}$ at a finite affine parameter. 
  
Finally, we can extend the vector field $\xi^{(p)}$ defined on the null geodesics $\gamma_{(p)}$ into a vector field $\xi$ defined on the entire $(d+1)$-dimensional manifold $\mM$ by requiring that
\eq{
& \xi \big|_{\gamma_{(p)}} = \xi^{(p)},  \label{s2:nullllinea} \\
& (\mathcal{L}_\xi g )_{\mu\nu}\sim O(r^{-\alpha_{\mu\nu }}) \quad \textrm{as } r\to \infty \label{s2:ask}, \\
& \xi \big|_{\gamma_\xi}=0, \label{s2:fixpointa}\\
& \nabla^\mu \xi^\nu \big|_{\gamma_\xi} = 2\pi n^{\mu\nu}.  \label{s2:xip1a}
}
The first condition~\eqref{s2:nullllinea} requires $\xi$ to be tangent to the null line $\gamma_{(p)}$ with a normalization that is fixed by \eqref{s2:nmbk} and the boundary condition \eqref{s2:xipbc}. Since $\gamma_{(p)}$ is null, its tangent vector is also normal to $\gamma_{(p)}$ and as a result of Frobenius' theorem we have
\eq{
\xi_{[\mu}\nabla_{\nu}\xi_{\lambda]}\big|_{\gamma_{(p)}}=0, \label{s2:Frobeniusa}
}
which is the generalization of eq.~\eqref{s2:xip2}. The second condition~\eqref{s2:ask}  requires $\xi$ to be an asymptotic Killing vector at the boundary ($r\to\infty$), which must be compatible with the boundary conditions $\delta g_{\mu\nu} \sim O(r^{-\alpha_{\mu\nu }})$ of the gravitational theory where $\alpha_{\mu\nu}$ are constants.  The third line~\eqref{s2:fixpointa} defines $\gamma_\xi$ as the set of fixed points of $\xi$, which naturally contains the set of fixed points of $\xi^{(p)}$, namely $\cup_{p\in\p\mA} \p\gamma_{(p)} \subset \gamma_\xi$. Finally, \eqref{s2:xip1a} implies that $\xi$ is the boost generator in the local Rindler frame near $\gamma_\xi$ such that the parameter $\tau$ defined by 
\eq{
2\pi \p_\tau = \xi,
 }
 corresponds to a local Rindler time coordinate on the two dimensional plane normal to $\gamma_\xi$ at each point in $\gamma_\xi$.
 
A vector field satisfying eqs.~\eqref{s2:nullllinea} --~\eqref{s2:xip1a} always exists, although it is not uniquely determined away from $\mA$, $\gamma_\xi$, and $\gamma_{(p)}$~\cite{Lashkari:2016idm}.  The non-uniqueness of $\xi$ away from these surfaces is not a problem, however, since we are only interested on the value of $\xi$ on the swing surface $\gamma_\mA$ and the boundary subregion $\mA$.
 

\subsection{The minimal swing surface proposal} \label{se:proposal}
In this section we extend the replica trick to the bulk and derive the holographic entanglement entropy following the lines of argument of Lewkowycz and Maldacena~\cite{Lewkowycz:2013nqa}. In what follows, we assume that the gravitational theory in the bulk is described by Einstein gravity. 

In section~\ref{se:localmodflow} we defined a vector field $\xi$ that is tangent to the set of null geodesics emanating from the boundary $\p\mA$ of subregion $\mA$ at the boundary and features a set of fixed points $\gamma_\xi$ in the bulk. A crucial difference between models of non-AdS holography and standard AdS/CFT is that in the former case, the boundary points $p\in\p\mA$ are not fixed points of the bulk modular flow generated by $\xi$, namely $\p\mA\not\subset\gamma_\xi$. This is also reflected in our continuation of the local Rindler time and the replica trick from the boundary to the bulk. In order to see this, let us consider a subset of $\gamma_\xi$ that is bounded by the null geodesics $\gamma_{(p)}$ such that
  \eq{
\gamma\subset\gamma_\xi, \qquad \p\gamma=\gamma_\xi\cap\gamma_{b\p} = \cup_{p\in\p \mA} \p \gamma_{(p)},
  }
where $\p\gamma$ is the boundary of $\gamma$ and $\gamma_{b\p} = \cup_{p\in\p\mA}\gamma_{(p)}$ is the collection of null geodesics~$\gamma_{(p)}$. The local Rindler time $\tau$ can be extended from the boundary endpoints $p\in\p \cal A$ to $\p\gamma_{(p)}$ along the null geodesics $\gamma_{(p)}$, which further continues along the bulk surface $\gamma$. This construction of the Rindler time function allows us to extend the replica trick into the bulk.

With this preparation, we can use the Lewkowycz-Maldacena and Dong-Lewkowycz-Rangamani arguments~\cite{Lewkowycz:2013nqa,Dong:2016hjy} to derive the bulk description of entanglement entropy. The derivation follows closely the steps of Lewkowycz and Maldacena~\cite{Lewkowycz:2013nqa} for states with time reversal symmetry which admit an Euclidean formulation. For more general states a similar argument along the lines of~\cite{Dong:2016hjy} is expected. In the following we will first determine the set of fixed points of the replica symmetry from the bulk equations of motion and then calculate the generalized gravitational entropy. 

Let us first extend the replica trick to the bulk. We will argue that $\gamma$ is the set of fixed points of the replica symmetry as $n\to1$ and that $\gamma$ is an extremal surface bounded by $\gamma_{b\p}$. The metric near the surface $\gamma$ can be expanded as
\eq{
ds^2=d\rho^2+\rho^2 d\tau_E^2+\big(h_{ij}+2x^a K_{aij} \big)dy^idy^j+\cdots,  \qquad \tau_E\sim \tau_E+2\pi, \label{s2:nh}
}
where $y^i$ denote the coordinates on the co-dimension two surface $\gamma$, $\rho$ denotes the distance from $\gamma$, $\tau_E$ is the Wick rotation of the local Rindler time $\tau$,  and $a$ stands for indices on the plane $(\rho,\tau_E)$ orthogonal to $\gamma$. The metric on $\gamma$ can be expanded in terms of $\rho$, with the linear order term given by the extrinsic curvature tensor $K_{aij}$ of $\gamma$. Note that all of the bulk fields including $K_{aij}$ are collectively denoted by $\phi$ and satisfy the periodic boundary conditions $\phi(\tau_E)=\phi(\tau _E + 2\pi )$.\footnote{There might be subtleties in this identification related to the presence of nonlocal boundary conditions, as discussed for example in~\cite{Song:2016pwx}.}
  
In analogy with the discussion in~\cite{Lewkowycz:2013nqa}, we can define the $n$-th smooth cover $\mM_n$ of the bulk manifold $\mM$ by changing the period of $\tau_E$ in \eqref{s2:nh} to $\tau_E\sim \tau_E+2\pi  n$. Furthermore, we can define the $Z_n$ quotient of $\mM_n$ by $\hat \mM_n =\mM_n/Z_n$, the latter of which features conical singularities at the loci of fixed points of $Z_n$ that are denoted by $\gamma_n$. In order to calculate the entanglement entropy we must take the $n\to1$ limit and our construction indicates that $\gamma_n \big|_{n\to1}$ is just the surface $\gamma$.\footnote{Here we have assumed that there are no additional saddles in the $n\to1$ limit.} The metric $\hat{g}_n$ of $\hat{\cal M}_n$ in the transverse directions can be  
expanded near $\gamma$ as 
 \eq{
  ds_{\hat g_n}^2=n^2 d\rho^2+\rho^2 d\tau_E^2+\cdots,\qquad \tau_E\sim \tau_E+2\pi .
  }
Note that assuming replica symmetry, the bulk fields on $\hat{\cal M}_n$ also satisfy the periodic boundary conditions $\phi(\tau_E)=\phi(\tau _E + 2\pi )$. Then the linearized equations of motion around $\gamma$ determine the shape of $\gamma$ when $n-1$ is small. In particular, Einstein's equations require the trace of the extrinsic curvature to vanish and hence $\gamma$ must be an extremal surface in Einstein gravity~\cite{Lewkowycz:2013nqa}. 

Let us now calculate the generalized gravitational entropy.  In order to accomplish this, we use the equivalence of the bulk and boundary partition functions and assume that the dominant saddle has replica symmetry in the bulk. The classical contribution to $\log Z_n$ in eq.~\eqref{EEbn} can be calculated in the bulk by the on-shell action $I[\cal M]$ on the $n$-th smooth cover $\mM_n$ of the original manifold $\mM$, 
  \eq{ 
	\log Z_n = -I[\mM_n]=-n I[{\hat \mM}_n].
  }
The entanglement entropy~\eqref{EEbn} can then be rewritten as
  \eq{
  S_{\mA}&=n^2 \p_n I[\hat g_n] \big|_{n\to 1}\underset{\text{on-shell}}{=} \int_{\p{\hat \mM}_n}   \bm \Theta[g,\p_n \hat g_n] \big|_{n=1}, \label{s2:SEEgamma0}  
  }
where we have omitted terms proportional to the bulk equations of motion which vanish on-shell and the presymplectic potential $\bm\Theta[g,\p_n \hat g_n]$ comes from total derivative terms in the variation of the action. In eq.~\eqref{s2:SEEgamma0},  $\p{\hat \mM}_n$ denotes all the co-dimension one boundaries of the manifold ${\hat \mM}_n$, which include the asymptotic boundary as well as the tip at $\rho\to 0$. By assumption, the boundary conditions guarantee that the asymptotic boundary does not contribute to the on-shell action. As a result, the only boundary contributions to \eqref{s2:SEEgamma0} come from the boundary at $\rho = 0$.\footnote{It is possible that there are corner contributions from the boundary $\p\gamma$ of $\gamma$ but we do not see such terms in the explicit examples discussed later in this paper.} For Einstein gravity, the entanglement entropy is thus given by
\eq{
S_{\mA}&=\int_{\gamma\times S^1}  \bm\Theta[g,\p_n \hat g_n] \big|_{n=1} =\int_{\gamma} \xi \cdot  \bm\Theta[g,\p_n \hat g_n] \big|_{n=1} \\
& = {{\rm Area} (\gamma)\over 4G}= {{\rm Area} (\gamma_{\mA})\over 4G}. \label{s2:SEEgamma}
}
Combining \eqref{s2:SEEgamma} together with the fact that $\gamma$ is an extremal surface bounded by the $\gamma_{b\p}$ surfaces, our proposal for holographic entanglement entropy is given by the minimal area of all the surfaces $X_{\mA}$ such that
\eq{
S_{\mA}= \min  \underset {X_{\cal A}\sim \cal A }{\textrm{ext}}  {{\rm Area}(X_{\cal A} )\over 4G},\qquad X_{\cal A} =X\cup \gamma_{b\p}\label{proposal},
}
where $X$ is a spacelike surface that hangs from $\p\mA$ by the collection of null ropes $\gamma_{b\p}$. All of the  surfaces $X_{\mA}$ are homologous to $\cal A$ which we denote by $X_{\cal A}\sim \mA$. The entanglement entropy is then given by the area of the swing surface $\gamma_{\cal A}=\gamma\cup \gamma_{b\p}$ where the bench $\gamma$  minimizes the area of the surfaces $X$ connecting the ropes $\gamma_{b\p}$.

The crucial difference between \eqref{proposal} and the RT/HRT proposal in AdS/CFT is that the surface $\gamma$ is not directly attached to the boundary at $\p\mA$ but is instead connected to $\p\mA$ by the null ropes $\gamma_{b\p}$. This guarantees that the swing surface is homologous to the subregion $\mA$ at the boundary. Note that in the derivation of \eqref{proposal} we have assumed that the bulk theory is described by Einstein gravity. In more general theories of gravity, we expect the area to be replaced by a generalized area which depends on the gravitational theory under consideration~\cite{Dong:2017xht}. In particular, in higher derivative theories of gravity the generalized area contains the Wald term and terms involving extrinsic curvatures~\cite{Dong:2013qoa}.

Note that since the bench is not directly anchored to $\partial \mA$, the holographic entanglement entropy could potentially be free of divergences. This implies that the entanglement entropy does not necessarily satisfy the area law at the boundary or, in two dimensions, that it does not have a logarithmic dependence on the length of the subsystem. We will see that in flat$_3$/BMSFT the entanglement entropy is indeed divergence-free and that it does not have any logarithmic dependence on the size of the subsystem. On the other hand, in (W)AdS$_3$/WCFT the entanglement entropy depends logarithmically on the size of the subsystem along one direction and it depends linearly on the size of the other direction. In this case such a linear dependence is related to the fact that WCFTs are nonlocal along one direction.

Let us conclude by summarizing our proposal for holographic entanglement entropy in holographic dualities beyond AdS/CFT satisfying the assumptions listed at the beginning of Section~\ref{se:generalproposal}. The holographic entanglement entropy of a subregion $\mA$ at the boundary is given by \eqref{proposal} and can be calculated by following these steps:
\begin{enumerate}
\item[$(1)$]  For each $p\in\p \mA$, find the approximate modular flow generator $\zeta^{(p)}$ as described in Section \ref{se:localmodflow}.
\item[$(2)$]   For each $p\in\p \mA$, find the null geodesic $\gamma_{(p)}$ emanating from $p$ whose tangent vector is an asymptotic Killing vector that reduces to $\zeta^{(p)}$ at the  asymptotic boundary.
\item[$(3)$]   Find the extremal surface spanning the region bounded by $\gamma_{b\p}=\cup_{p \in \p\mA} \gamma_{(p)}$.
\item[$(4)$]  If there are multiple extremal surfaces choose the one with the minimal area.
\end{enumerate}
In these steps, one needs to be careful about the cutoff at infinity, as we will see in more detail in the following sections.


\subsection{Relative entropy for general states}

In this section we derive general expressions for the relative entropy of general states and provide a formal check of our proposal~\eqref{proposal} in Einstein gravity. Relative entropy is a quantum information measure that quantifies the distinguishability between two states $\rho$ and $\sigma$ that is defined by
\eq{
S(\rho||\sigma) \equiv \tr(\rho \log \rho)-\tr(\rho \log \sigma)  = \Delta \EV{{\cal H}_{mod}(\sigma)}-\Delta S_{\mA}. \label{s2:relativeentropy}
}
In \eqref{s2:relativeentropy}, $\Delta \EV{{\cal H}_{mod} (\sigma) } = -\tr(\rho \log \sigma) + \tr(\sigma \log \sigma)$ is the difference in the expectation value of the modular Hamiltonian associated with the state $\sigma$ while $\Delta S_{\mA} = - \tr(\rho \log \rho) + \tr(\sigma \log \sigma)$ is the difference of the entanglement entropy. Relative entropy has several important features including: ($i$) it is positive definite, namely $S(\rho||\sigma) \ge 0$; and (${ii}$) it monotonically increases with the size of the system such that $S(\rho_\mA||\sigma_\mA) \le S(\rho_{\mathcal B}||\sigma_{\mathcal B} )$ where $\mA$ and $\mathcal B$ are two subregions satisfying $\mA\subset \mathcal B$.

Let us now consider the holographic dual of relative entropy between the vacuum state $\sigma$ and an excited state $\rho$. For the vacuum state we assume that the exact modular flow generator is known in both the boundary and the bulk, which we respectively denote by $\zeta_{vac}$ and $\xi_{vac}$. On the other hand, for the excited state there is no exact Killing vector generating the modular flow. Nevertheless, as discussed in Section~\ref{se:localmodflow}, we can always find a vector field that behaves like the exact modular flow generator in the vacuum state as it approaches $\gamma_\mA$. This vector field is denoted by $\xi$ and satisfies eqs.~\eqref{s2:ask} ---~\eqref{s2:xip1a} together with the condition 
\eq{
\xi \big|_{\p \mM} =\zeta_{vac}. \label{bc1}
}

The approximate modular flow generator $\xi$ can be used to define the quasilocal energy in the $d$-dimensional non-timelike region $\Sigma$ that is bounded by $\cal A$ at the boundary and the swing surface $\gamma_{\cal A}$ in the bulk such that $\p \Sigma =  \cal A \cup \gamma_{\cal A}$. Using the covariant phase space formalism~\cite{Wald:1993nt,Iyer:1994ys,Barnich:2001jy},  the diffeomorphism associated with $\xi$ is generated by the Hamiltonian $\delta {\mathscr H}_\xi[\phi ]$
\eq{
\delta {\mathscr H}_\xi^{\Sigma}[\phi ] =\int_{\Sigma} d{\bm \chi}_{\xi}[\phi, \delta \phi], \label{s2:xicharge}
}
 where the $d$-form $d{\bm \chi}_{\xi}[\phi, \delta \phi]$ is determined by the gravitational action, $\phi$ collectively denotes the bulk fields, and $\delta \phi$ denotes their on-shell variations. If there exists a $(d-1)$-form $\bm Q_\xi[\phi] $ and  a $d$-form ${\bm K[\phi]}$ such that ${\bm \chi}_{\xi}[\phi, \delta \phi] = \delta(\bm Q_\xi[\phi] - \xi\cdot \bm K[\phi])$ on $\p\Sigma$, then the gravitational charge in the region $\Sigma$ is integrable and can be written as a surface charge on $\p\Sigma= \cal A \cup \gamma_{\cal A}$,
\eq{
{ \mathscr H}_\xi^{\Sigma} [\phi] =  \int_{\p\Sigma}  \big(   \bm Q_\xi[\phi] - \xi\cdot \bm K[\phi]   \big) \equiv {\cal Q}_\xi^{\p\Sigma}= {\cal Q}_\xi^{\mA}- {\cal Q}_\xi^{\gamma_{\mA}},\label{s2:Wald}
}
where we split the integral over the asymptotic boundary $\mA$ and the swing surface $\gamma_{\mA}$. 

We now show that the proposed holographic entanglement entropy~\eqref{proposal} is given by the surface charge  ${\cal Q}_\xi^{\gamma_\mA}={\cal Q}_\xi^{\gamma}+{\cal Q}_\xi^{\gamma_{b\p}}$  evaluated along the swing surface in Einstein gravity, where
$\bm Q_\xi[g]$ is given by
\eq{
{\bm Q}_\xi[g] =-\frac{1}{16\pi G(d-1) ! } \nabla^\mu \xi^\nu \epsilon_{\mu\nu \mu_3 \dots \mu_{d+1}} dx^{\mu_3}\wedge \dots \wedge dx^{\mu_{d+1}}.
}
Let us first consider the contribution of the null ropes $\gamma_{b\p}$ to the surface charge ${\cal Q}_\xi^{\gamma_\mA}$. Since the vector field $\xi$ is orthogonal to $\gamma_{b\p}$, Frobenius' condition~\eqref{s2:Frobeniusa} implies that $\int_{\gamma_{b\p}} {\bm Q}_\xi[g] \propto \int_{\gamma_{b\p}} \xi^{[\alpha}\nabla^\mu\xi^{\nu]} d\lambda \wedge\cdots =0$. Furthermore, the fact that $\xi$ is tangent to $\gamma_{b\p}$ and $\bm K[\phi]$ is antisymmetric means that  $\int_{\gamma_{b\p}} \xi\cdot {\bm K}[\phi]=0$ as well. Consequently, the charge on the  ropes vanishes. Let us now consider the surface charge ${\cal Q}_\xi^{\gamma}$ evaluated on the bench $\gamma$ of the swing surface. In this case, the fixed point condition~\eqref{s2:fixpointa} implies that $\int_{\gamma} \xi\cdot \bm K[g]  = 0$ while the normalization~\eqref{s2:xip1a} implies that ${\bm Q}_\xi[g]$ is proportional to the area of the bench. As a result, we find that the surface charge along the swing surface is given by the area of the bench in Planck units,
\eq{
{\cal Q}_\xi^{\gamma_\mA} =  {\text{Area}(\gamma)\over 4G} =  {\text{Area}(\gamma_\mA)\over 4G}= S_{\mA},\label{SwaldSA}
}
which is the holographic entanglement entropy according to our proposal~\eqref{proposal}.

On the other hand, as demonstrated in~\cite{Apolo:2020qjm}, the surface charge evaluated along $\mA$ is holographically dual to the expectation value of the  modular Hamiltonian ${\cal H}_{mod}$ between two states,
\eq{
\Delta \langle {\cal H}_{mod} \rangle = \Delta {\cal Q}_\xi^{\mA}.\label{bkmodular}
}
Using the holographic dictionary \eqref{SwaldSA} and \eqref{bkmodular}, we can define the holographic dual of relative entropy, in analogy with the discussion in AdS/CFT~\cite{Lashkari:2016idm},
\eq{
S \big( \rho_\mA||\sigma_\mA \big)=\Delta \langle {\cal H}_{mod} \rangle - \Delta S_{\mA} = { \mathscr H}_\xi^{\Sigma} [\phi] - { \mathscr H}_\xi^{\Sigma} [\phi_{vac}] . 
}
In particular, when $\rho$ is an infinitesimal perturbation above the vacuum $\sigma$, $\xi$ is an exact Killing vector of the background fields $\phi$ and $\delta{\mathscr H}^\Sigma_\xi[\phi] = \delta {\mathcal Q}_\xi^{\mA}[\phi]- \delta {\mathcal Q}_\xi^{\gamma_\mA}[\phi]=0$ implies the first law of entanglement entropy. In contrast, when $\rho$ is a finite perturbation of the vacuum state $\sigma$, $\xi$ is no longer an exact Killing vector and the relative entropy can be non-zero. The modular Hamiltonian in explicit models of non-AdS holography is studied in detail in the companion paper~\cite{Apolo:2020qjm}. 


\subsection{Remarks}

In this section we comment on a few consequences of our proposal for holographic entanglement entropy, including the entanglement entropy for multiple intervals, deformations of the swing surface, and the entanglement wedge.

\subsubsection*{Multiple intervals}

For three-dimensional bulk spacetimes, the proposal~\eqref{proposal} implies that the entanglement entropy for multiple intervals can be obtained by pairing the endpoints and summing over the areas of the swing surfaces associated with each   pair. Generically, there are several different ways of pairing points that is consistent with the homology constraint. In this case, as in the RT/HRT prescription in AdS/CFT, we must choose the configuration that yields the minimal area. 

\begin{figure}[h] 
   \centering
    \includegraphics[scale=0.611]{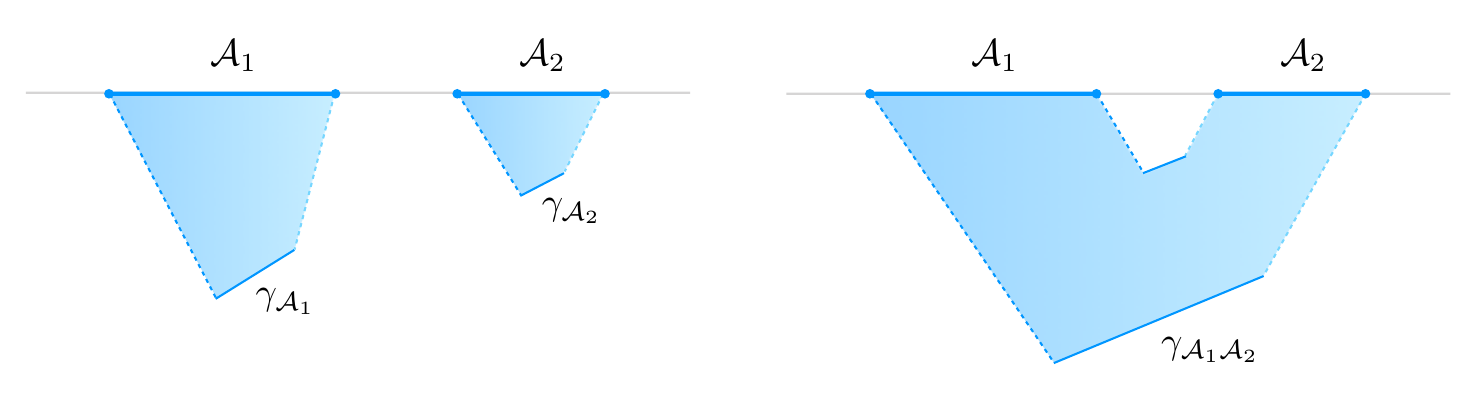} 
    \caption{Two-dimensional sketch of two competing swing surfaces $\gamma_{\mA_1} \cup \gamma_{\mA_2}$ and $\gamma_{\mA_1 \mA_2}$ homologous to the disjoint union of two intervals $\mA_1$ and $\mA_2$. The dotted lines denote the ropes while the solid lines denote the benches. 
    }
\label{s2:multipleinterval} 
\end{figure}

As an example, let us consider the disjoint union of two intervals $\mA_1$ and $\mA_2$. Then, the holographic entanglement entropy is given in terms of the area by
  \eq{
  S_{\mA_1 \cup \mA_2} =\frac{1}{4G} \textrm{min} \Big( {\rm Area}(\gamma_{\mA_1} \cup \gamma_{\mA_2}),\, {\rm Area}(\gamma_{\mA_1 \mA_2}) \Big),
  }
  where $\gamma_{\mA_1} \cup \gamma_{\mA_2}$ and $\gamma_{\mA_1 \mA_2}$ denote the two possible swing surfaces depicted in Fig.~\ref{s2:multipleinterval}.  In particular, for two intervals that are far apart from each other we clearly have
  \eq{
  S_{\mA_1 \cup \mA_2} =\frac{ {\rm Area}(\gamma_{\mA_1} \cup \gamma_{\mA_2})}{4G},
  }
and the mutual information $I = S_{\mA_1} + S_{\mA_2} - S_{\mA_1 \cup \mA_2}$ vanishes. A phase transition occurs when the intervals $\mA_1$ and $\mA_2$ are close enough to each other such that ${\rm Area}(\gamma_{\mA_1} \cup \gamma_{\mA_2}) = {\rm Area}(\gamma_{\mA_1 \mA_2})$ and the entropy is given by 
  \eq{
  S_{\mA_1 \cup \mA_2} =\frac{ {\rm Area}(\gamma_{\mA_1 \mA_2})}{4G}.
  }
This observation is consistent with the calculation of mutual information of WCFTs in~\cite{Chen:2019xpb}. Mutual and 3-partite information have also been discussed in flat holography in~\cite{Asadi:2018lzr}.

%


\subsubsection*{Towards the entanglement wedge}

Let us comment briefly on the entanglement wedge associated with the swing surface. Let $y^i$ with $i=1,\dots, d-1$ denote the coordinates parametrizing the co-dimension two surface $\gamma_\xi$ in the bulk. Starting from any point on $\gamma_\xi$ there is an  outgoing, future-directed null geodesic orthogonal to $\gamma_\xi$. We denote the vector tangent to this geodesic by $l^\mu\p_\mu=\p_\lambda$ with $\lambda$ the affine parameter whose value on $\gamma_\xi$ is chosen to be zero.

The collection of such null geodesics span $N_+$, the future directed part of the light-sheet of $\gamma_\xi$ with non-positive expansion. By requiring that the coordinates $y^i$ remain the same along the null geodesics on $N_+$, all points on $N_+$ can be parametrized by $(\lambda, y^i)$. Following the discussion in Appendix~\ref{app:nullgeodesics}, we can then find a vector field $\xi $ satisfying $\xi^\nu\nabla_\nu \xi^\mu|_{N_+}=2\pi$ by requiring that 
\eq{
\xi^\mu= 2\pi (d\lambda/ d\tau) l^\mu,
}
where $(d^2 \tau/ d\lambda^2) = - ({d\tau/ d\lambda})^2$. The solution to this equation is $\tau=\tau_0 + \log| \lambda|$ where $\tau_0$ corresponds  to the freedom of rescaling the affine parameter. From this construction, it can be explicitly seen that $\xi|_{\gamma_\xi}=0$, which tells us that $\gamma_\xi$ is the set of fixed points of the vector field $\xi$.  Similarly, we can define the past directed part of the light-sheet $N_-$ and define $\xi$ on it as well. By construction, all the null ropes satisfy $\gamma_{(p)} \subset N_+\cup N_-$. Consistency of the holographic duality also requires $(N_+\cup N_-)\cap \p \cal M=\p\mathcal{D}$ where $\p\cal D$ is the boundary of the causal domain $\cal D$ of $\mA$. A \emph{pre-entanglement wedge} $W_{\xi}$ can be defined as the region bounded by $N_+\cup N_-\cup \cal D$. We expect the entanglement wedge to be a subset of the pre-entanglement wedge as the former should be associated with $\gamma_{\cal A}$ instead of $\gamma_\xi$. We leave this and related questions for further work.
%
\begin{figure}[ht!] 
   \centering
    \includegraphics[scale=0.6875]{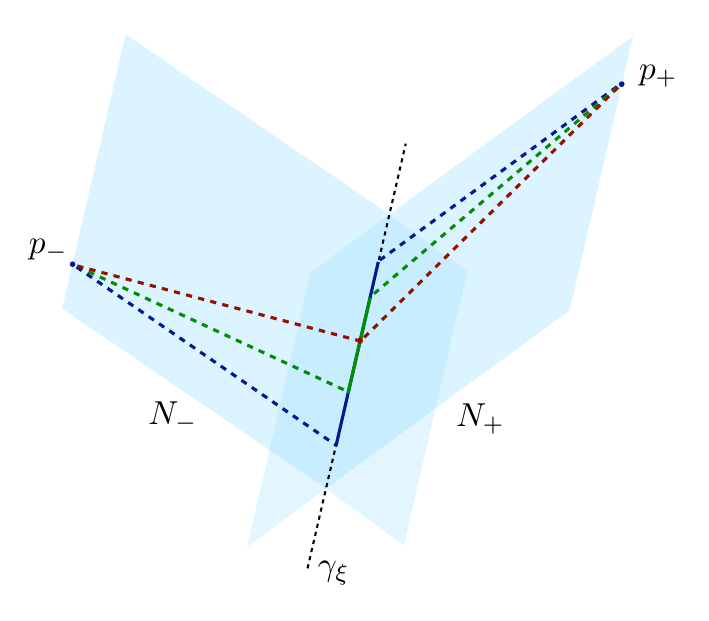} 
    \caption{Deformations (green and red) of the swing surface (blue) in $N_+\cup N_-$. The dashed lines are ropes and the solid lines are benches. The ropes of the original swing surface are null while the deformed ones are spacelike. Note that the bench of the red surface has shrunk to a point.  Such deformations preserve the area when the expansion $\theta$ on $N_+\cup N_-$ vanishes and decrease the area when $\theta$ is negative.  }
\label{s2:shrunkswingsurface} 
\end{figure}
%
\subsubsection*{Deformations of the swing surface}
Let us now consider deforming the swing surface $\gamma_{\mathcal A}$ along $N_+$. For convenience we work with $d = 2$. Consider an infinitesimal deformation along the null geodesics such that the change of the area $A$ is related to the expansion by $\lim_{A\to0}{1\over A}{dA\over d\lambda}= \theta $. In the examples where an explicit Rindler transformation is known, $\xi$ corresponds to an exact Killing vector in the bulk and the light-sheets $N_{\pm}$ are Killing horizons with zero expansion. In these examples, we can deform one of the ropes of the swing surface, for example $\gamma_+$, to a spacelike geodesic $\gamma_+' \subset N_+$ whose intersection with $\gamma_\xi$ remains on the original bench $\gamma$. In this case, the bench $\gamma'$ of the deformed swing surface is a subset of $\gamma$. Nevertheless, the total length of the deformed swing surface $\gamma_{\mA}' = \gamma_+'\cup \gamma'\cup\gamma_-$ remains the same. We can similarly deform the rope $\gamma_- \to \gamma_-'$ within $N_-$. In particular, we can shrink the entire bench to a point such that the deformed swing surface becomes $\gamma_{\mathcal A}=\gamma_+'\cup \gamma_-'$, as illustrated by the red curve in Fig.~\ref{s2:shrunkswingsurface}. This phenomenon has also been observed in the context of flat holography~\cite{Hijano:2017eii}.  For more general backgrounds, a deformation of the swing surface within the lightsheets $N_+\cup N_-$ will generically decrease its length if the expansion $\theta$ is negative. This is similar to the property of the HRT surface in the AdS/CFT correspondence \cite{Hubeny:2007xt}.

To summarize, we conclude that a deformation of the swing surface within the lightsheets $N_+\cup N_-$ does not increase the total area. We expect that this property of swing surfaces can be used to prove useful entanglement entropy inequalities, such as strong subadditivity \cite{Wall:2012uf}. However, since the swing surface contains null ropes which cannot be included in a spacelike surface, the maximin argument used in \cite{Wall:2012uf} can not be directly adopted. In this paper, we perform a few explicit checks of strong subadditivity in both flat$_3$/BMSFT and AdS$_3$/WCFTs on zero mode backgrounds. In these examples, we find that strong subadditivity is satisfied in WCFTs, while it can be violated in  BMSFTs.  We leave a general discussion of entanglement entropy inequalities for future work. For related discussions on other inequalities of entanglement entropy in BMSFTs and WCFTs see \cite{Grumiller:2019xna,Detournay:2020vrd}.


\section{Swing surfaces in flat$_3$/BMSFT} \label{se:flat}

In this section we describe how the general proposal for holographic entanglement entropy works for asymptotically flat spacetimes in the flat$_3$/BMSFT correspondence. More concretely, we derive the swing surface for zero-mode backgrounds, compute the variation of the entanglement entropy that results from perturbations of these spacetimes, and discuss strong subadditivity in these models of non-AdS holography.

Einstein gravity in asymptotically flat three-dimensional spacetimes admits consistent boundary conditions at future null infinity that lead to a set of asymptotic symmetries described by the three-dimensional BMS algebra~\cite{Barnich:2006av}. This observation leads to the conjectured duality between gravity in asymptotically flat spacetimes and BMS-invariant field theories at the boundary, namely to the flat$_3$/BMSFT correspondence~\cite{Barnich:2010eb,Bagchi:2010eg,Bagchi:2012cy}. In the bulk, the space of solutions of pure Einstein gravity in Bondi gauge is described by
\eq{
ds^2 =  \Theta(\phi) du^2-2 du dr+2 \Big[\Xi(\phi)+\frac12 u\partial_\phi\Theta(\phi) \Big] du d\phi+r^2 d\phi^2, \label{flatbackground}
}
where $\phi \sim \phi + 2 \pi$ and $\Theta(\phi)$, $\Xi(\phi)$ are two periodic functions. 
When $\Theta(\phi) = M$ and $2\Xi(\phi) = J$ are constants, the metric~\eqref{flatbackground} describes zero-mode backgrounds with energy ${M/8G}$ and angular momentum ${J/8G}$. These solutions include, in particular, the global Minkowski vacuum with $M = -1$ and $J = 0$, conical defect geometries where $-1 < M < 0$, and flat cosmological solutions where $M > 0$.

Entanglement entropy in BMSFTs has been computed for single intervals in the vacuum as well as thermal and more general states in~\cite{Bagchi:2014iea,Basu:2015evh,Hosseini:2015uba,Jiang:2017ecm,Grumiller:2019xna}. In particular, ref.~\cite{Jiang:2017ecm} found that the geometric dual of entanglement entropy in flat spacetimes is not given by the RT/HRT surface, but instead by what we call a swing surface in this paper (see Fig.~\ref{s2:swingsurface}). The latter motivates our proposal~\eqref{proposal}. The results of~\cite{Jiang:2017ecm} are based on the generalized Rindler method. This method exploits the symmetries of BMSFTs and the vacuum state to devise a BMS transformation that maps the causal domain of dependence $\cal D$ of an interval $\mA$ on the original manifold $\p \mM$ to a generalized Rindler spacetime. Since this is a symmetry transformation, the entanglement entropy associated with the interval $\mA \in \p\mM$ on the vacuum can be obtained from the thermal entropy in Rindler space. This result can be generalized to thermal states by an additional symmetry transformation. 

For zero-mode backgrounds, the generalized Rindler method can be extended into the bulk where it can be used to determine the swing surface whose area reproduces the entanglement entropy of the BMSFT at the boundary. If we parametrize the interval $\mA$ along the $u$ and $\phi$ coordinates at the boundary by $l_u$ and $l_\phi$ respectively, then the holographic entanglement entropy is given by
  \eq{
  S_{\mA} = \frac{1}{4G} \bigg[ \sqrt{M} \Big( l_u+\frac{J l_\phi}{2M}\Big) \coth\Big(\frac{\sqrt{M}l_\phi}{2}\Big)-\frac{J}{M}  \bigg]. \label{flatEE}
  }
Using the holographic dictionary, the entropy~\eqref{flatEE} matches the entanglement entropy of BMSFTs on thermal states~\cite{Bagchi:2014iea,Jiang:2017ecm}. In what follows, we will show that the proposal~\eqref{proposal} reproduces the entanglement entropy of zero-mode backgrounds given in \eqref{flatEE} without using the Rindler method. We will also consider linearized fluctuations around zero-mode backgrounds and show that the variation of the entanglement entropy matches the variation of the modular Hamiltonian.


\subsection{Approximate modular flow} \label{se:flatlocalmodflow}

The modular flow generator for the vacuum state in BMSFTs was first written down in~\cite{Jiang:2017ecm} and has been studied in more detail in~\cite{Hijano:2017eii, Godet:2019wje,Fareghbal:2019czx,Apolo:2020qjm}. As described in Section~\ref{se:generalproposal}, this modular flow generator plays an important role in the derivation of holographic entanglement entropy for both the vacuum and more general states. In this section we write down the approximate modular flow generator for general states near the endpoints of the interval at the boundary, which we then use to determine the ropes of the swing surface.

Let us consider an interval $\mA \in \p\mM$ whose endpoints $\p\mA$ are given in terms of $(u,\phi)$ coordinates by 
\eq{ 
\p\mA = \big\{ \big(  u_-,\phi_- \big), \, \big(  u_+,\phi_+ \big)\big\}, \qquad u_+ - u_-= l_u, \quad \phi_+-\phi_-=l_\phi, \label{interval}
}
where $l_u$ and $l_\phi$ are assumed to be positive. The approximate modular flow generator for general states can be obtained near the endpoints~\eqref{interval} in the following way. Consider first the vacuum state on the plane. The modular flow generator $\zeta$ is the symmetry generator~\eqref{s2:modflowboundary} that leaves the causal domain $\cal D$ of the interval $\mA$ invariant. Its normalization is fixed by requiring that $e^{i\zeta}$ maps points in $\cal D$ back to themselves. These conditions allow us to determine each of the $a_i$ coefficients featured in \eqref{s2:modflowboundary}, as has been explicitly done in~\cite{Apolo:2020qjm}, with the result
\eq{
  \!\!\! {\zeta \over 2\pi} & =  \bigg[ { u_+ (\phi-\phi_-)^2-u_- (\phi-\phi_+)^2\over(\phi_--\phi_+)^2} + {u (2 \phi-\phi_--\phi_+)\over \phi_--\phi_+} \bigg] \p_u + {\left(\phi-\phi_-\right) \left(\phi-\phi_+\right)\over\phi_--\phi_+} \p_\phi . \label{generalzeta}
}
The approximate modular flow generator $\zeta^{(p)}$ for general states around the endpoint $p \in \p\mA$ can be obtained from this expression by sending the other endpoint of $\p\mA$ to infinity. We consequently find
\eq{
\zeta^{(\pm)} = \mp 2\pi \big[ \big(u-u_\pm \big) \p_u + \big(\phi-\phi_\pm\big) \p_\phi \big], \label{localmodflow}
}
where $\zeta^{(\pm)}$ denotes the approximate modular flow generator associated with the $(u_\pm, \phi_\pm)$ endpoint. In particular, note that the endpoints $(u_\pm, \phi_\pm)$ are fixed points of the approximate modular flow generator, a property inherited from the exact modular flow generator of the vacuum state.

The ropes of the swing surface correspond to null geodesics $\gamma_\pm$ that emanate from the endpoints $(u_\pm, \phi_\pm)$ of the boundary interval $\mA$. As described in Section~\ref{se:proposal}, the tangent vector of these geodesics must reduce to the approximate modular flow generator~\eqref{localmodflow} at a cutoff surface near the asymptotic boundary. Since $\zeta^{(\pm)}$ vanishes at the endpoints of $\mA$, the tangent vectors of $\gamma_\pm$ are parallel to the radial direction $\p_r $ at the cutoff surface. On the other hand, it is not difficult to verify that any vector that points along the radial direction is null everywhere in spacetimes that are solutions of pure Einstein gravity in the Bondi gauge~\eqref{flatbackground}. Therefore,  the ropes $\gamma_{\pm}$ of the swing surface can be described in the region $r \in [0, \infty)$ by $u=u_\pm, \phi=\phi_\pm$.  There are subtleties in the parametrization of these geodesics in the Bondi gauge beyond $r = 0$, as discussed in detail in Appendix~\ref{app:zeroMode2}. For the zero-mode solutions~\eqref{flatbackground} with $M>0$, the range of the radial coordinate $r$ can be extended to negative values, such that the ropes $\gamma_{\pm}$ of the swing surface lie along radial null geodesics that are given by 
\eq{ 
\gamma_\pm: \; u=u_\pm, \;\phi= \phi_\pm, \quad r \in (-\infty, \infty). \label{flatropesgeneral}
} 
The parametrization of the ropes for zero-mode backgrounds with $M<0$ can be found in Appendix~\ref{app:swingzeromode2}.

We have shown that the ropes $\gamma_\pm$ emanating from the endpoints $(u_\pm, \phi_\pm)$ of the interval at the boundary are given by null geodesics along the radial coordinate of the asymptotically flat spacetimes~\eqref{flatbackground}. According to our general proposal~\eqref{proposal}, the holographic entanglement entropy is then given by the   area of the extremal surface in the bulk that connects the null geodesics $\gamma_+$ and $\gamma_-$, as described in detail next.


\subsection{The Poincar\'{e} vacuum} 

Before considering more general cases, let us briefly illustrate how the propoosal \eqref{proposal} works for single intervals on the Poincar\'e vacuum.  The Poincar\'e vacuum describes the vacuum state of the dual BMSFT on the plane and its metric is simply given by
\eq{
ds^2= -2 dudr +r^2 d\zc^2, \label{poincare} 
}
where $\zc \in (-\infty, \infty)$. For convenience, we consider a symmetric interval at the boundary whose endpoints $\p \mA$ are parametrized in terms of the $(u,z)$ coordinates by 
\eq{ 
\p\mA = \Big\{ \Big(  -\frac{l_u}{2}-,-\frac{l_\zc}{2} \Big), \, \Big(  \frac{l_u}{2},\frac{l_\zc}{2} \Big)\Big\}. \label{syminterval}
}
where $l_u$ and $l_z$ are both positive.

As discussed in the previous section, the null geodesics $\gamma_{\pm}$ emanating from the endpoints~\eqref{syminterval} are tangent to $\p_r$ and given by
  \eq{
  \gamma_\pm: \, u=\pm \frac{l_u}{2}, \quad z=\pm \frac{l_\zc}{2}, \quad r \in (-\infty, \infty). \label{flatropespoincare}
  }
The next step in the holographic entropy proposal described in Section~\ref{se:proposal} consists of finding the extremal surface $X$ that lies between the null geodesics~\eqref{flatropespoincare}. In flat space this surface is just a line --- a spacelike geodesic to be more precise --- that connects $\gamma_+$ to $\gamma_-$. The length of the surfaces $X_{\mA} = \gamma_- \cup X \cup \gamma_+$ is thus given 
  \eq{
  L(r_+, r_-) =  \sqrt{r_- r_+ l_\zc^2 + 2 (r_- - r_+)l_u}, \label{poincarelength}
  }
where, $r_\pm$ denote the points along $\gamma_\pm$ where $X$ is attached. The extremum of \eqref{poincarelength} is found at
\eq{
r_{-} = - r_{+} =  \frac{2l_u}{ l_\zc^2}, \label{lambdapmflat}
}
which determines the location of the swing surface $\gamma_\mA = \gamma_- \cup \gamma \cup \gamma_+$. In particular, the bench $\gamma$ of the swing surface can be parametrized in terms of the $\zc$ coordinate by
\eq{
\gamma:\, u(\zc) &= \frac{l_u l_\zc}{8\zc} + \frac{l_u \zc}{2 l_\zc}, \quad r(\zc) = -\frac{l_u}{\zc l_\zc}, \label{flatX}
}
where $\zc \in (-\infty, -l_\zc/2] \cup [ l_\zc/2, \infty)$.\footnote{There is a subtlety with the range of $\zc$ in the parametrization of the bench \eqref{flatX}, as discussed in detail in Appendix~\ref{app:benchcommentsPoincare}.} Finally,  the area of the swing surface $\gamma_{\mA}$ is given by
  \eq{
  S_{\cal A} = \frac{\textrm{Area}(\gamma_{\cal A})}{4 G} = \frac{l_u}{2G l_\zc}. \label{flat: poincareholoEE}
  }
This expression agrees with the entanglement entropy computed on the field theory side of the flat$_3$/BMSFT correspondence, which can be seen by taking the $J \to 0$ and $M \to 0$ limits in \eqref{flatEE}.


\subsection{Zero-mode backgrounds}  \label{se:zeromode}

Let us now test the holographic entropy proposal~\eqref{proposal} on the more general zero-mode backgrounds describing flat cosmological solutions with $M > 0$
\eq{
ds^{2} =M du^{2} -2 dudr +J du d\phi +r^{2} d\phi^{2},  \qquad \phi \sim \phi + 2 \pi. \label{zeroMode}
}
These backgrounds can be regarded as the flat limit of the BTZ black hole in AdS$_3$~\cite{Barnich:2012aw}. The case $M < 0$, which includes the global Minkowski vacuum, is also interesting and is treated in detail in Appendix~\ref{app:swingzeromode2}. For convenience, we consider once again a symmetric interval $\mA$ at the boundary that is given by \eqref{interval} with
\eq{
(u_{\pm},\phi_\pm) = \Big(\pm\frac{l_u}{2}, \pm \frac{l_\phi}{2} \Big). \label{syminterval2}
}

 According to our prescription~\eqref{proposal}, the holographic entanglement entropy is given by the distance of a geodesic connecting two points on the ropes $\gamma_+$ and $\gamma_-$, which are the radial null geodesics described in \eqref{flatropesgeneral}. The distance between these geodesics is given by
  \eq{
  \begin{split}
 M L(r_+, r_-)^2 &=
 l_u^2 M^2+2 l_u M \big(l_\phi \sqrt{M} r_c + r_- - r_+\big)+r_c^2 \big(l_\phi^2 M+2\big)\\
 & - 2 \cosh \big(l_\phi \sqrt{M}\big) \big(r_c^2 - r_- r_+ \big) +2 l_\phi \sqrt{M} r_c (r_- - r_+) \\
 &+2 r_c \big(r_+ - r_- \big) \sinh \big(l_\phi \sqrt{M}\big)-2 r_- r_+,
 \end{split}\label{Ldistformula}
 }
 where we used $r_c \equiv J/2\sqrt{M}$ while $r_+$ and $r_-$ denote the radial coordinates of two points along $\gamma_+$ and $\gamma_-$, respectively. We find that the extremum of \eqref{Ldistformula} corresponds to the following choice of $r_{\pm}$
   \eq{
    r_- =  - r_+= \frac{l_u M+l_\phi \sqrt{M} r_c-r_c \sinh \big(l_\phi \sqrt{M}\big)}{\cosh \big(l_\phi \sqrt{M}\big)-1}.
   }
These values of $r_\pm$ determine the points along the ropes $\gamma_\pm$  \eqref{flatropesgeneral}  where the bench $\gamma$ of the swing surface is attached (see Appendix~\ref{app:benchcomments} for the parametrization of the bench). The holographic entanglement entropy is then given by the area of the swing surface $\gamma_\mA = \gamma_- \cup \gamma \cup \gamma_+$, namely
\eq{
S_{\mA}  &= \frac{\textrm{Area}(\gamma_\mA)}{4G} = \frac{1}{4G} \bigg| \sqrt{M} \Big( l_u+\frac{J l_\phi}{2M}\Big) \coth\Big(\frac{\sqrt{M}l_\phi}{2}\Big)-\frac{J}{M}  \bigg|. \label{flathee}
}
The absolute value guarantees that the entanglement entropy, which corresponds to the length of the swing surface, is a positive quantity. When the argument of the absolute value is positive, eq.~\eqref{flathee} agrees with the entanglement entropy of BMSFTs at finite temperature in~\eqref{flatEE}. It would be interesting to understand how the absolute value in the bulk computation of the entanglement entropy emerges in the field theory side of the calculation, a question that we leave for future work.
 

\subsection{Beyond zero-mode backgrounds}

In this subsection we apply the general prescription for holographic entanglement entropy to arbitrary linearized fluctuations of the zero-mode background~\eqref{zeroMode}  and show that the first law of entanglement entropy is satisfied in flat$_3$/BMSFT.

Let us consider linearized perturbations of the metric compatible with the parametrization of the phase space in \eqref{flatbackground}, namely
  \eq{
  { \delta g}_{\mu\nu} dx^{\mu} dx^{\nu} = {\delta \Theta}(\phi) du^{2} + \Big[ {\delta \Xi}(\phi) + \frac{1}{2} u \p_{\phi} {\delta \Theta}(\phi) \Big] du d\phi. \label{perturbation}
  }
Note that the parametrization of the null geodesics $\gamma_{\pm}$ in \eqref{flatropesgeneral} is independent of the parameters of the background and remain null in the perturbed spacetime. Also, since the bench of the swing surface corresponds to an extremal configuration, it remains extremal to linear order in the perturbations of the metric. As a result, the only effect of the linearized perturbation~\eqref{perturbation} is to change the area (length) of the swing surface. The variation of the length of the spacelike bench is thus given by
  \eq{
  \delta L(r_+, r_-) & = \int_{\gamma} d \phi \, \delta \sqrt{{g}_{\mu\nu} \p_{\phi} x^{\mu} \p_{\phi}x^{\nu}}.
  }
 Using the parametrization of the bench given in eqs.~\eqref{bondibenchr} and~\eqref{bondibenchu} we find that  $\delta L(r_+, r_-)$ can be conveniently written as
  \eq{
  \delta L(r_+, r_-) & =  \frac{1}{4\pi} \int_{-l_\phi/2}^{l_\phi/2}  d\phi \big[{T}(\phi)\delta\Theta(\phi) + 2{Y}(\phi) \delta\Xi(\phi)\big],
  }
where, as discussed in Appendix~\ref{app:benchcomments}, we have deformed the contour of integration so that $\phi \in [-l_\phi /2, l_\phi/2]$, while the ${T}(\phi)$ and ${Y}(\phi)$ functions read
  \eq{
  \begin{split}
  \!\!{T} (\phi) &= \frac{\pi}{2 M \sinh\Big(\tfrac{\sqrt{M} l_{\phi}}{2}\Big)} \bigg\{  (J l_\phi + 2 l_u M) \bigg[ \textrm{coth}\Big(\tfrac{\sqrt{M} l_{\phi}}{2}\Big) \cosh(\sqrt{M}\phi)- \textrm{csch}\Big(\tfrac{\sqrt{M} l_{\phi}}{2}\Big)   \bigg]   \\ 
  &  \hspace{.5cm} + \frac{2 J}{\sqrt{M}} \bigg[ \cosh(\sqrt{M}\phi) -  \cosh \Big(\tfrac{\sqrt{M} l_{\phi}}{2}\Big)\bigg]- 2J \phi \sinh(\sqrt{M}\phi) \bigg\},
  \end{split} \\
  \!\! {Y} (\phi) &= \frac{2\pi}{\sqrt{M}\sinh\Big(\tfrac{\sqrt{M} l_{\phi}}{2}\Big)}\Big[ \cosh\Big(\tfrac{\sqrt{M} l_{\phi}}{2}\Big) - \cosh(\sqrt{M}\phi)\Big]. 
  }
The variation of the holographic entanglement entropy is thus given by
\eq{
 \delta S_{\mA}= \frac{ \delta L(r_+,r_-)}{4G}=\frac{1}{16\pi G}\int_{-l_\phi/2}^{l_\phi/ 2}  d\phi \big[{T}(\phi)\delta\Theta(\phi)+2{Y}(\phi) \delta\Xi(\phi)\big]. \label{flatlinear}
}
It is not difficult to verify that for zero-mode fluctuations where $\delta \Theta(\phi) = \delta M$ and $\delta \Xi(\phi) = \delta J/2$, eq.~\eqref{flatlinear} can be integrated within the space of solutions, which reproduces the finite expression for the entanglement entropy given in \eqref{flathee}. 
 
In Section~\ref{se:generalproposal}, we argued that the area of the swing surface agrees with the gravitational charge associated with the modular flow generator evaluated along the swing surface. This can be explicitly shown for perturbations of zero-mode backgrounds in flat$_3$/BMSFT. Indeed, it is not difficult to show that the infinitesimal gravitational charge $\delta {\cal Q}_\xi^{\gamma_{\cal A}}[g]$ on the swing surface receives no contributions from  the ropes and that the result agrees with the variation of the area \eqref{flatlinear}, namely
  \eq{
 \delta {\cal Q}_\xi^{\gamma_{\cal A}}[g] =\delta {\cal Q}_\xi^{\gamma}[g] = \delta S_{\mA}.\label{deltaSapp}
      }
In addition, we note that \eqref{flatlinear} also matches the infinitesimal charge $\delta {\cal Q}_\xi^{{\cal A}}[g]$ evaluated on the boundary interval $\cal A$. This charge corresponds to the variation of the modular Hamiltonian, as shown explicitly in Section 3 of~\cite{Apolo:2020qjm}. We thus find that the swing surface proposal satisfies the first law of entanglement entropy, 
  \eq{
  \delta S_{\mA} = \delta {\cal Q}_\xi^{\gamma_{\cal A}}[g]= \delta {\cal Q}_\xi^{{\cal A}}[g] = \delta \Vev{{\cal H}_{mod}},
  }
in agreement with the general discussion in Section~\ref{se:proposal}.

To conclude, we have calculated the holographic entanglement entropy in flat$_3$/BMSFT using the swing surface proposal~\eqref{proposal} on zero-mode backgrounds and linearized fluctuations of these spacetimes, and checked explicitly that the first law of entanglement entropy holds. These results provide additional evidence that swing surfaces are the geometric duals of entanglement entropy  in BMS-invariant field theories.


\subsection{Comments on strong subadditivity}
Let us now comment on the implications of entanglement entropy for single intervals to strong subadditivity. For adjacent intervals $\cal A$, $\cal B$, and $\cal C$ at the boundary,  strong subadditivity is the statement that 
\eq{\label{SSA}
S_{\cal A B} + S_{\cal B C} \geq S_{\cal B}+S_{\cal A B C},
}
where $\cal A B \equiv \mathcal A \cup \mathcal B$. We consider the following cases.

\subsubsection{Fixed slope}

For simplicity, let us  first consider three intervals with the same slope. In these cases, the entanglement entropy of the interval $\mA$ at the boundary is a function of a length scale $L$, namely $S_\mA = S(L)$. Thus, in order to test if strong subadditivity is satisfied, we need to test if the entanglement entropy is upper convex (i.e. concave), namely if 
\eq{
S''(L) \leq 0.
} 
In what follows we will test which of the zero mode backgrounds described in this section satisfy strong subadditivity. In particular, we will show that for intervals with the same slope strong subadditivity is satisfied for the Minkowski and Poincar\'e vacua but that it can fail for backgrounds with conical deficits and for flat cosmological solutions.

\vspace{5pt}
 
\noindent{\bf Global Minkowski vacuum.} For the global Minkowski vacuum given in \eqref{flatbackground} with $ \Theta(\phi) = M = -1$ and $2\Xi(\phi)  = J = 0$, the entanglement entropy \eqref{app:globalminkEE} satisfies
\eq{
S''(l_\phi)  = \frac{l_u}{4 G l_\phi}  \Big[ \frac{l_\phi}{2}\cot \Big( \frac{l_\phi}{2}\Big) -1 \Big] \csc^2 \Big( \frac{l_\phi}{2}\Big)\le0.
}
Since $S(l_\phi)$ is upper convex, we find that strong subadditivity is satisfied for adjacent intervals with fixed slope ${l_u/ l_\phi}$.

\vspace{5pt}

\noindent{\bf Poincar\'e vacuum.} For the Poincar\'e vacuum \eqref{poincare}, the holographic entanglement entropy \eqref{flat: poincareholoEE} is constant and proportional to the slope ${l_u/ l_z}$ of the interval at the boundary. As a result, the entanglement entropy saturates strong subadditivity, i.e.~$S''(l_z) = 0$.

\vspace{5pt}

\noindent{\bf Conical defects.} The conical defect geometries are described by \eqref{flatbackground} with $ \Theta(\phi) = M$ and $-1 < M < 0$. Their entanglement entropy is derived in \eqref{app:conicaldefectHEE} and its second derivative with respect to $l_\phi$ is given by
\eq{
S''(l_\phi)  = -s \frac{M}{4 G} \Big({l_u\over l_\phi} + \frac{J}{2M}\Big)  \Big[ \frac{\sqrt{-M} l_\phi}{2}\cot \Big( \frac{ \sqrt{-M}l_\phi}{2}\Big) -1 \Big] \csc^2 \Big( \frac{ \sqrt{-M} l_\phi}{2}\Big) \le 0,
}
where $s=\text{sign}\Big[ \sqrt{-M} \Big( l_u + \tfrac{J  l_{\phi}}{2M} \Big) \cot\Big(\tfrac{\sqrt{-M}  l_{\phi}}{2}\Big) - \tfrac{J}{M}\Big]$ takes into account the effect of the absolute value in~\eqref{app:conicaldefectHEE}.  As a result, we find that $S''(l_\phi) \le 0$ implies
\eq{
\bigg[  \Big( \frac{l_u}{l_\phi} + \frac{J  }{2M}  \Big) + \frac{J}{(-M)^{3/2}  l_\phi}\tan\Big(\frac{\sqrt{-M}  l_{\phi}}{2}\Big) \bigg]   \Big({l_u\over l_\phi} + \frac{J}{2M}\Big) \ge 0.
}
Therefore, strong subadditivity is satisfied on conical defect geometries with positive angular momentum $J>0$  for intervals with $l_u>0$, $l_\phi>0$, and ${l_u\over l_\phi}+{J\over 2M}>0$.

\vspace{5pt}

\noindent{\bf Flat cosmological solutions.} For flat cosmological solutions we have $\Theta(\phi) = M > 0$ in eq.~\eqref{flatbackground} and the entanglement entropy \eqref{flathee} satisfies
\eq{
S''(l_\phi)  = s\frac{M}{4 G} \Big({l_u\over l_\phi} + \frac{J}{2M}\Big)  \Big[ \frac{\sqrt{M} l_\phi}{2}\coth \Big( \frac{ \sqrt{M}l_\phi}{2}\Big) -1 \Big] \textrm{csch}^2 \Big( \frac{ \sqrt{M} l_\phi}{2}\Big),
}
where  $s=\text{sign} \Big[ \sqrt{M} \Big( l_u+\tfrac{J l_\phi}{2M}\Big) \coth\Big(\tfrac{\sqrt{M}l_\phi}{2}\Big)-\tfrac{J}{M} \Big]$. Since $M$ is positive, requiring $S''(l_\phi) \le 0$ gives rise to the condition 
\eq{
\bigg[    \Big(\frac{ l_u}{ l_\phi}+\frac{J }{2M}\Big) -\frac{J}{M^{3/2} l_\phi } \tanh\Big(\frac{\sqrt{M}l_\phi}{2}\Big) \bigg] \Big({l_u\over l_\phi} + \frac{J}{2M}
\Big)   \le 0.
}
Hence, strong subadditivity is always violated for intervals with $l_u>0$ and $l_\phi>0$ on flat cosmological solutions with positive angular momentum $J>0$.


\subsubsection{Changing the slope}

Thus far we have only considered intervals with the same slope and found that strong subadditivity can be violated in flat cosmological solutions. We now show that for more general intervals, strong subadditivity can also be violated in other backgrounds. In order to illustrate this let us consider the Poincar\'e vacuum and three adjacent intervals $\mA$, $\mB$, $\mC$  parametrized by $(l_u, l_z) = (d_\mA,l_\mA)$, $(d_\mB, l_\mB)$, $(d_\mC,l_\mC)$, all of which are assumed to be positive. This means that the interval $\mA\mB$ is given by $(d_\mA + d_\mB, l_\mA + l_\mB)$ and similarly for the intervals $\mB\mC$ and $\mA\mB\mC$. Strong subadditivity~\eqref{SSA} then implies  
\eq{
\frac{d_\mA+d_\mB}{l_\mA+l_\mB}+\frac{d_\mB+d_\mC}{l_\mB+l_\mC} \ge
\frac{ d_\mB}{ l_\mB}+\frac{d_\mA+d_\mB+d_\mC}{ l_\mA+l_\mB+l_\mC},
}
which leads to the condition
\eq{
(d_\mA +  d_\mB) \Big(\frac{d_\mB}{l_\mB} - \frac{d_\mC }{l_\mC}\Big) +(d_\mB+d_\mC)\Big(\frac{d_\mB}{l_\mB} - \frac{d_ \mA}{l_\mA}\Big) \le 0 .
}
We thus find that strong subadditivity can be violated in the Poincar\'e vacuum for some configurations of the intervals. A sufficient but not necessary condition for strong subadditivity to be satisfied is for the slope of the middle interval $\mB$ to be smaller or equal than the slopes of the two adjacent intervals $\mA$ and $\mC$, namely that $\frac{d_\mB}{l_\mB}\le \frac{d_\mA}{l_\mA},\frac{d_\mC }{l_\mC}$.

\vspace{5pt}

To summarize, we find that the fate of strong subadditivity of entanglement entropy in BMSFT depends on the background as well as the subregions, in contrast to AdS/CFT where strong subadditivity is universally satisfied. It would be interesting to understand the origin of the violations of strong subadditivity in flat holography which we leave for future work.


\section{Swing surfaces in (W)AdS$_3$/WCFT} \label{se:warped}

In this section we use the holographic entanglement entropy proposal~\eqref{proposal} to construct the swing surfaces of zero-mode asymptotically AdS$_3$ backgrounds satisfying Dirichlet-Neumann boundary conditions~\cite{Compere:2013bya}. These swing surfaces agree with the geometric picture put forward in~\cite{Song:2016gtd} and reproduce the entanglement entropy of single intervals in WCFTs at the boundary~\cite{Castro:2015csg,Song:2016gtd}. Note that the backgrounds considered in this section are also compatible with the Brown-Henneaux boundary conditions~\cite{Brown:1986nw} that lead to a dual CFT at the boundary. We will show that in this case the bench of the swing surface is pushed all the way to the boundary and reduces to the standard RT/HRT surface~\cite{Ryu:2006bv,Hubeny:2007xt}. Finally, we will consider warped AdS$_3$ backgrounds and show that our results apply equally well to this class of spacetimes.

Three-dimensional Einstein gravity with a negative cosmological constant admits alternative boundary conditions from those of Brown and Henneaux~\cite{Brown:1986nw} that lead to different asymptotic symmetries at the boundary~\cite{Compere:2008us,Compere:2013bya,Troessaert:2013fma,Avery:2013dja,Apolo:2014tua}. In particular, the asymptotic symmetries compatible with the Dirichlet-Neumann CSS boundary conditions are described by a Virasoro-Kac-Moody algebra~\cite{Compere:2013bya}. These are the same symmetries characterizing a class of nonrelativistic quantum field theories known as warped CFTs~\cite{Hofman:2011zj, Detournay:2012pc}, a fact that motivates the so-called AdS$_3$/WCFT correspondence. The space of solutions of Einstein gravity satisfying CSS boundary conditions can be described in Fefferman-Graham gauge by two functions $L(u)$ and $J(u)$, as well as a constant $T_v$, such that
\eq{
\begin{split}
    ds^{2} &=  {dr^{2} \over r^{2}} + r^{2} du \big[ dv + J'(u) du \big ] +  L(u) (du)^{2} +  T_v^2 \big [ dv +  J'(u) du \big ]^{2} \\
    & \hspace{.5cm} + {1 \over r^{2}} T_v^2 L (u) du \Big [ dv +  J'(u) du \Big ],  \label{wcft:adsmetric}
  \end{split}
}
 where we have set the radius of AdS$_3$ to one and the lightcone coordinates $(u,v)$ satisfy $(u,v) \sim (u + 2\pi, v+ 2\pi)$. The latter are related to the coordinates $(x,y)$ of the WCFT at the boundary at $r\to \infty$ by a state-dependent map~\cite{Detournay:2012pc} 
  \eq{
  u = x, \qquad v  = x + \frac{y}{2\mu T_v}, \label{wcft:map}
  }
where $\mu$ is a parameter characterizing the WCFT. 

In analogy with the asymptotically flat spacetimes discussed in Section~\ref{se:flat}, it is helpful to discuss the zero-mode backgrounds described by \eqref{wcft:adsmetric}. When ${L}(u) = T_u^2$ and $J(u) $ are constants, the backgrounds~\eqref{wcft:adsmetric} reduce to zero-mode solutions of three-dimensional gravity that are also compatible with Brown-Henneaux boundary conditions, and whose energy and angular momentum are respectively given by ${\cal E} = (T_u^2 +T_v^2)/4G$ and ${\cal J} = (T_u^2 - T_v^2)/4G$. The zero-mode backgrounds described by \eqref{wcft:adsmetric} include the global AdS$_3$ vacuum when ${\cal E} = -1/8G$ and ${\cal J} = 0$, conical defect geometries when $-1/8G < {\cal E} < 0$, and BTZ black holes when ${\cal E} \ge |{\cal J}| \ge 0$.

The generalized Rindler method described briefly in Section~\ref{se:flat} works in a similar way in the (W)AdS$_3$/WCFT correspondence~\cite{Castro:2015csg,Song:2016gtd}. In particular, the holographic entanglement entropy associated with an interval $\mA$ at the boundary is reproduced by the area of a swing surface in the bulk. For zero-mode backgrounds, the holographic entanglement entropy is given by~\cite{Song:2016gtd}
   \eq{
    S_{\cal A} = \frac{T_v l_v}{4G} +\frac{1}{4G} \log \bigg[ \frac{ \sinh (T_u l_u)} { \varepsilon_{} T_u }\bigg], \label{wcft:eezero}
   }
where $l_u$ and $l_v$ denote the lengths of the interval $\mA$ along the $u$ and $v$ coordinates, and $\varepsilon$ is the UV cutoff in the WCFT. Note that the gravitational $T_u$ and $T_v$ parameters are related to the thermodynamic potentials of the WCFT at the boundary. Using the holographic dictionary, one finds that \eqref{wcft:eezero} reproduces the entanglement entropy for single intervals in thermal states in WCFTs. In the following we will show how \eqref{wcft:eezero} can be obtained from the general prescription for holographic entanglement entropy proposed in Section~\ref{se:proposal}. In addition, we will discuss the role played by the boundary conditions in the derivation of the swing surface and describe how the swing surface reduces to the RT/HRT surface in AdS/CFT.

\subsubsection*{Comments on strong subadditivity}
It is interesting to note that the entanglement entropy for zero mode backgrounds \eqref{wcft:eezero} satisfies strong subadditivity for adjacent intervals. Indeed, we note that the first term of \eqref{wcft:eezero} is linear in $l_v$ and does not contribute to the check of strong subadditivity, while the second term is upper convex in $l_u$, namely it satisfies
\eq{
S''(l_u) = - \frac{T_u^2}{4G} \textrm{csch}^2(T_u l_u)\le 0,
}
 for both real and imaginary values of $T_u$. This means that strong subadditivity is satisfied for the global AdS$_3$ vacuum, the Poincar\'e vacuum, conical defect geometries, and BTZ black holes. 
This result is to be expected since the second term in \eqref{wcft:eezero} is the same as that of a chiral half of a two-dimensional CFT, which is known to satisfy strong subadditivity. 
 

\subsection{Approximate modular flow} \label{se:bulkprelude}

The first step in our proposal for holographic entanglement entropy is the identification of the approximate modular flow generator near the endpoints $\p\mA$ of the interval $\mA$ at the boundary. In this section we will write down the approximate modular flow generator for general states which we will use in the next section to construct the ropes $\gamma_\pm$ of the swing surface.

Let us consider a WCFT on the \emph{canonical cylinder} where the Virasoro and $U(1)$ coordinates, denoted by $x$ and $y$ respectively, satisfy $(x,y) \sim (x + 2\pi ,y)$. On the canonical cylinder, the vacuum expectation values of zero-mode charges depend on a parameter $\mu$ that is given in holographic WCFTs by $\mu = \sqrt{-c/6k}$, where $c$ is the central charge and $k$ is the $U(1)$ level of the Virasoro-Kac-Moody algebra. In particular, these vacuum expectation values can be removed by a warped conformal transformation that takes the canonical cylinder to the so-called {\it reference cylinder} where $(x,y) \sim (x + 2\pi, y - i \mu)$ \cite{Chen:2019xpb}. We consider a general parametrization of the interval $\mA$ on the canonical cylinder such that its endpoints are given by 
\eq{
\p \mA = \big\{ \big(x_-, y_- \big), \big(x_+, y_+ \big) \big \}, \qquad x_+ - x_- = l_x, \quad y_+ - y_- = l_y. 
\label{wcft:interval}
}
The modular flow generator on the vacuum of such a WCFT can be obtained via the Rindler method. It can also be derived by finding the linear combination of the vacuum symmetry generators that leaves the causal domain of dependence of $\mA$ invariant. For a general interval, the exact modular flow generator on the vacuum state satisfies~\cite{Apolo:2020qjm}
  \eq{
  \zeta =  \frac{i }{e^{2\pi i (x_+ - x_-)} - 1} \Big [ e^{2\pi i (x -x_-)}  + e^{-2\pi i (x -x_+)} - e^{2\pi i (x_+ - x_-)} - 1 \Big] \p_x -2\pi \mu \p_y. \label{wcft:modflow}
  }
We can obtain the approximate modular flow $\zeta^{(p)}$ for general states near the endpoint $p \in \p\mA$ by sending the other endpoint of $\mA$ to infinity. The approximate modular flow generator is thus given by
\eq{
\zeta^{(\pm)} = \mp 2\pi (x - x_{\pm} ) \p_x  - 2\pi\mu \p_y, \label{wcft:approxmodflow}
}
where the $\pm$  superscript refers to the $(x_\pm, y_\pm)$ endpoint.

It is instructive to compare \eqref{wcft:approxmodflow} to the approximate modular flow of a CFT. The latter can be obtained in a similar way from the exact modular flow generator on the vacuum state found in e.g.~\cite{Czech:2019vih}. If we denote the two chiral coordinates of the CFT by $u$ and $v$, the approximate modular flow generator is then given by
  \eq{
  \zeta^{(\pm)}_{CFT}  = \mp [ 2\pi (u -u_{\pm} ) \p_u  -  2\pi (v - v_{\pm}) \p_v],\label{wcft:approxmodflowcft}
  }
which corresponds to the boost generator near each of the endpoints $(u_\pm, v_\pm)$. A crucial difference between eqs.~\eqref{wcft:approxmodflow} and \eqref{wcft:approxmodflowcft} is the finite $2\pi\mu \p_y$ term characteristic of holographic WCFTs. This term guarantees that $\zeta^{(\pm)}$ does not vanish at the endpoints of the interval $\mA$, in contrast with the behavior of $\zeta^{(\pm)}_{CFT}$. Furthermore, we see that up to this term, the approximate modular flow generator~\eqref{wcft:approxmodflow} corresponds to half of that of a CFT, which follows from the fact that WCFTs feature only one copy of the Virasoro algebra. 

We note that the endpoints $\p\mA$ of the interval $\mA$ are fixed points of the modular flow generator~\eqref{wcft:approxmodflow} despite the presence of the $2\pi\mu\p_y$ term. As argued in~\cite{Chen:2019xpb}, the replica trick in WCFT must be understood as opening up the local reference plane near each of the endpoints and gluing circles satisfying the identification $(z,y)\sim (z e^{2\pi i},y-i\mu)$. The $2\pi\mu \p_y$ term guarantees that $\zeta^{(p)}$ is compatible with the replica trick and that it maps the endpoints of $\mA$ back to themselves. Furthermore, as shown in Appendix~\ref{app:aproxmodflowwarped}, the $2\pi\mu \p_y$ term is necessary to extend the approximate modular flow generator $\zeta^{(p)}$ into an asymptotic Killing vector in the bulk, the latter of which played an important role in the derivation of the swing surface proposal in Section~\ref{se:proposal}. Finally, we note that the $2\pi\mu \p_y$ term is also necessary to match the entanglement entropy in the bulk and boundary sides of the AdS$_3$/WCFT correspondence~\cite{Song:2016gtd}. 

The next step in the holographic entropy proposal is to determine the null geodesics $\gamma_\pm$ extending from the endpoints $\p\mA$ at the boundary into the interior of the spacetime. Note that since the vector tangent to the ropes must reduce to~\eqref{wcft:approxmodflow} at the boundary, the former must have components along $\p_y = (1/2 \mu T_v) \p_v$ where we used the map~\eqref{wcft:map} relating the WCFT and AdS$_3$ lightcone coordinates. Due to the dependence on $T_v$, we expect the parametrization of the ropes to depend on the background spacetime, as we explicitly verify in the next section. Once the ropes $\gamma_\pm$ are determined, the swing surface can be obtained by finding the extremal surface connecting $\gamma_+$ and $\gamma_-$. 


\subsection{Swing surfaces for zero-mode backgrounds} \label{se:ssads}
We now test the holographic entanglement entropy proposal and determine the swing surface of generic zero-mode backgrounds. For convenience, we work in the following gauge
  \eq{
    ds^2 &= {d\rho^2\over 4(\rho^2-4T_u^2T_v^2)} + \rho\, du dv+T_u^2 du^2+T_v^2 dv^2,  \label{wcft:zeromode}
  }
which is related to the Fefferman-Graham gauge used in \eqref{wcft:adsmetric} by the change of coordinates $\rho = r^2 + {T_u^2 T_v^2}/{r^2}$. In these coordinates the endpoints of the interval ${\cal A}$ at the boundary are parametrized by
  \eq{
  \p {\cal A} = \big\{ (u_-, v_-), (u_+,v_+) \big\}, \qquad u_+ - u_- = l_u, \quad v_+ - v_- = l_v, \label{wcft:intervalbulk}
  }
and are related to the field theory parametrization used in \eqref{wcft:interval} via the map~\eqref{wcft:map}.

Let us first determine the ropes $\gamma_\pm$ of the swing surface. As described in Section~\ref{se:proposal}, the ropes $\gamma_\pm$ are null geodesics emanating from the boundary endpoints \eqref{wcft:intervalbulk} whose tangent vectors reduce to the approximate modular flow generator \eqref{wcft:approxmodflow} at the boundary. Since the zero-mode backgrounds~\eqref{wcft:zeromode} feature two commuting Killing vectors $\p_u$ and $\p_v$, the null geodesics $\gamma_\pm$ satisfy two conservation equations
  \eq{
 T_u^2 u' + {\rho v'\over2} = p_u, \qquad T_v^2 v' + {\rho u'\over2} = -p_v, \label{wcft:geodeq1}
  }
  where primes denote derivatives with respect to the affine parameter $\lambda$ while $p_u$ and $-p_v$ denote the momenta along the $u$ and $v$ coordinates.  Furthermore, since the ropes of the swing surface are required to be null, we also have
\eq{
\frac{\rho'^2}{4(\rho^2 - 4T_u^2 T_v^2)} + T_u^2 u'^2 + T_v^2 v'^2 + \rho \,u' v' = 0. \label{wcft:geodeq2} 
}

In Appendix~\ref{app:geodesicsads} we show that the solution to eqs.~\eqref{wcft:geodeq1} and~\eqref{wcft:geodeq2} that reaches the boundary as $\lambda\to \infty$ must necessarily have $p_u p_v\ge0$ and, furthermore, that null geodesics with $p_u p_v > 0$ do not satisfy the boundary condition \eqref{s2:xipbc}. Instead, the ropes of the swing surface can be obtained from the solutions to eqs.~\eqref{wcft:geodeq1} and~\eqref{wcft:geodeq2} with $p_v = 0$, which can be written as\footnote{The solutions with $p_u=0$ and $p_v\neq0$ are similar to eq.~\eqref{wcft:wcftsols} with $u \leftrightarrow v$. These geodesics have nonvanishing $\xi^{(\pm)u}$ components at the boundary and do not satisfy the boundary condition $\xi^{(\pm)}\big|_{\p \cal M}=\zeta^{(\pm)}$.}
  \renewcommand{\arraystretch}{.5}
 \eq{
   \gamma_\pm:  \, \left\{ \begin{array}{l}
    \rho = 4 |p_u|T_v \lambda + \rho_0 , \\
  \\
  u = \mp \frac{1}{4 T_u} \log\bigg(\dfrac{\rho+ 2 T_u T_v}{\rho - 2 T_u T_v} \bigg) + \tilde{u}_\pm, \\
  \\
  v = \mp \frac{1}{4 T_v} \log\big[( \rho^2 - 4 T_u^2 T_v^2)/\rho^2_\infty \big]  + \tilde{v}_{\pm},
  \end{array}\right. \label{wcft:wcftsols}
  }
  where $\rho_0$, $\tilde{u}_\pm$, and $\tilde{v}_\pm$ are integration constants, $\lambda$ is the affine parameter, and $\rho_\infty$ is a radial cutoff in the bulk. We have parametrized the solutions~\eqref{wcft:wcftsols} such that the $\gamma_\pm$ geodesic carries $\pm |p_u|$ momentum. Nevertheless, note that the value of $p_u$ can be absorbed by a rescaling of the affine parameter $\lambda$ that does not change the solution. It is also useful to note that at the cutoff surface $\rho = \rho_\infty$, the solutions \eqref{wcft:wcftsols} reach the point
\eq{
(u, v, \rho) = \bigg( \tilde{u}_\pm \mp \frac{T_v}{\rho_\infty}+{\cal O}\bigg(\frac{1}{\rho_\infty^2 } \bigg), \,  \tilde{v}_{\pm}  \mp \frac{T_u^2T_v}{\rho_\infty^2}+{\cal O}\bigg(\frac{1}{\rho_\infty^3 } \bigg),  \, \rho_\infty \bigg). \label{wcft:bulkendpoints}
}

We now show that the boundary condition \eqref{s2:xipbc} is satisfied by the null geodesics \eqref{wcft:wcftsols} and determine the values of the integration constants $\tilde u$, $\tilde v$, and $\rho_0$. As shown in Appendix~\ref{app:nullgeodesics}, we can always find a parameter $\tau(\lambda)$ such that the vector $\xi^{(\pm)}$ tangent to $\gamma_\pm$ satisfies the normalization condition \eqref{s2:nmbk}. As a result, the tangent vector of \eqref{wcft:wcftsols} can be written as
 \eqsp{
  \xi^{(\pm)} & \equiv 
  2\pi \frac{d\lambda}{d\tau}\frac{d x^{\mu}}{d\lambda}\p_{\mu} =\pm 2\pi \lambda (u' \p_u + v' \p_v + \rho' \p_\rho)  \\
 & =  \frac{2\pi T_v (\rho - \rho_0)}{\rho^2 - 4 T_u^2 T_v^2} \bigg( \p_u - \frac{\rho}{2 T_v^2} \p_v \bigg) \pm 2\pi (\rho - \rho_0) \p_\rho. \label{wcft:xi2}
  }
On the other hand, using the map \eqref{wcft:map}, we can rewrite the boundary modular flow generator \eqref{wcft:approxmodflow} in terms of the bulk coordinates such that
\eq{
\zeta^{(\pm)} = \mp2\pi (u - u_{\pm} ) (\p_u + \p_v) - \frac{\pi}{T_v} \p_v.\label{wcft:approxmodflowbulk} 
}
 Comparing the $u$ and $v$ components of the tangent vector \eqref{wcft:xi2} to the components of the approximate modular flow generator~\eqref{wcft:approxmodflowbulk}, we find that the boundary condition $\xi^{(\pm)}|_{\p\mM}  = \zeta^{(\pm)}$ is satisfied at the cutoff surface $\rho = \rho_\infty$ provided that $\rho_0 = 2T_v^2$ and
 \eq{
u = u_\pm \mp \frac{T_v}{\rho_\infty}+{\cal O}\bigg({1\over \rho_\infty^2 } \bigg). \label{wcft:uupm}  
}
Furthermore, comparing \eqref{wcft:uupm} to the point \eqref{wcft:bulkendpoints} reached by the null geodesics $\gamma_\pm$ at the cutoff surface, the later of which is also required to match \eqref{wcft:intervalbulk} as $\rho_\infty \to \infty$, we obtain
 \eq{ 
 \tilde{u}_\pm= u_\pm+{\cal O}\bigg({1\over \rho_\infty^2 } \bigg),\quad  \tilde{v}_\pm= v_\pm + {\cal O}\bigg({1\over \rho_\infty^2} \bigg) .
 } 
Consequently, the null geodesics with $p_v = 0$ given in \eqref{wcft:wcftsols} describe the ropes of the swing surface.

The next step in the holographic entropy proposal~\eqref{proposal} consists of finding the extremal surface lying between the ropes. The latter can be determined by extremizing the geodesic distance between two points on $\gamma_+$ and $\gamma_-$. In order to facilitate the discussion on warped AdS$_3$ backgrounds in Section~\ref{se:wads}, we derive the geodesic distance $L(p_1, p_2)$ between two spatially separated points $p_1 = (u_1, v_1, \rho_1)$ and $p_2 = (u_2, v_2, \rho_2)$ by exploiting the symmetries of locally AdS$_3$ spacetimes. The geodesic distance must be invariant under the simultaneous action of the local isometries of the zero-mode backgrounds~\eqref{wcft:zeromode}, namely under $SL(2,R)_L \times SL(2,R)_R$ transformations.\footnote{The $SL(2,R)_L$ generators are given by the following Killing vectors 
\eq{
 L_0 = -\frac{1}{2T_u} \p_u, \qquad L_{\pm 1} =e^{\pm 2 T_u u} \bigg[ -\frac{\rho}{T_u \sqrt{\rho^2 - 4 T_u^2 T_v^2}}\p_u + \frac{T_u}{\sqrt{\rho^2 - 4 T_u^2 T_v^2}}\p_v \pm    \sqrt{\rho^2 - 4 T_u^2 T_v^2} \p_\rho \bigg], \notag
 }  \label{wcft:sl2}
while the $SL(2,R)_R$ generators can be obtained from the expressions above by letting $u \leftrightarrow v$ and $T_u \leftrightarrow T_v$.
 }
  We first note that there are two independent functions $D_+(p_1, p_2)$ and $D_-(p_1, p_2)$ that are invariant under an $SL(2,R)_L \times U(1)_R$ subgroup where $U(1)_R$ generates translations along $v$. These functions are given by
 \eq{
   D_\pm = & \frac{e^{\pm T_v v_{12}}}{8 T_u T_v} \Big[e^{\pm T_u u_{12}} \sqrt{\big(\rho_1 + 2 T_u T_v \big) \big(\rho_2 + 2 T_u T_v \big)}- e^{\mp T_u u_{12}} \sqrt{\big(\rho_1 - 2 T_u T_v \big) \big(\rho_2 - 2 T_u T_v \big)} \, \Big], \label{wcft:Ddef}
  }
where $x_{12}^\mu\equiv x^\mu_1 - x^\mu_2$ for any two coordinates $x^\mu_1$ and $x^\mu_2$. Requiring $L(p_1, p_2)$ to be invariant under the full $SL(2,R)_R$ group implies that the geodesic distance depends on $p_1$ and $p_2$ only through the combination $D(p_1, p_2) = D_+ (p_1, p_2) + D_-(p_1, p_2)$. Furthermore, the functional $L(D)$ can be determined from the normalization condition $g^{\mu\nu} \p_\mu L(p_1,p_2) \p_\nu L(p_1,p_2) = 1$ where the derivative is taken with respect to one of the points, say $p_2$. As a result, we find that the geodesic distance $L(p_1, p_2)$ between two spatially separated points in the locally AdS$_3$ backgrounds~\eqref{wcft:zeromode} is given by
  \eq{
  L(p_1, p_2) = \cosh^{-1} \big[ D(p_1, p_2) \big]. \label{wcft:Ldef} 
  }

The location of the extremal surface lying between the $\gamma_+$ and $\gamma_-$ ropes is determined by extremizing the geodesic distance~\eqref{wcft:Ldef} where we take $p_1\in \gamma_-$ and $p_2\in \gamma_+$ with 
 \eq{
  \rho_1 = 2 T_u T_v\, \coth [ 2 T_u ( u_1 - u_-)], \qquad v_1= v_- -\frac{1}{2T_v} \log \Big\{ \frac{ \sinh [2 T_u ( u_1 - u_-)]}{2T_u T_v \rho_\infty^{-1}}\Big\}, \label{wcft:gammaminus} \\
  \rho_2 = 2 T_u T_v\, \coth [ 2 T_u ( u_+ - u_2)],  \qquad v_2 = v_+ + \frac{1}{2T_v} \log \Big\{ \frac{ \sinh[2 T_u ( u_+ - u_2)]}{2T_u T_v \rho_\infty^{-1}}\Big\}. \label{wcft:gammaplus}
  }
 In this parametrization, the $D_+(u_1, u_2)$ and $D_-(u_1, u_2)$ functions are given by
  \eq{
  D_+(u_1, u_2) &= e^{-l_v T_v} \rho_\infty^{-1} T_u T_v \, \textrm{csch} (2 T_u  u_{+1} )  \,\textrm{csch} (2 T_u u_{2-}) \sinh \big[T_u (l_u + 2 u_{21}) \big],  \label{wcft:Dplus} \\
    D_-(u_1, u_2)  &= \frac{e^{l_v T_v} \sinh (l_u T_u) }{4 \rho_\infty^{-1} T_u T_v}, \label{wcft:Dminus}
  }
where we note that $D_-(u_1, u_2)$ is independent of the $u_1$ and $u_2$ parameters. Solving the extremality conditions corresponds to solving $\p_{u_1} D_+ = \p_{u_2} D_+= 0$. In this way, we find that the endpoints of the bench are located at 
\eq{ 
u_1=u_2= \frac{u_+ + u_-}{2}. \label{wcft:intersection}  
}
The spacelike geodesic lying between the null ropes at the points determined by~\eqref{wcft:intersection} is a line with fixed values of $u$ and $\rho$ that extends along the $v$ direction
 \eq{
   \gamma: \, u=\frac{u_+ + u_-}{2} ,\quad 
   v \in\Big [\frac{v_+ + v_- - \Delta v}{2},\, \frac{v_+ + v_- + \Delta v}{2} \Big], \quad \rho = 2 T_u T_v \coth(T_u l_u), \label{wcft:gamma}
   } 
where $\Delta v$ is given by
  \eq{
   \Delta v \equiv l_v  + \frac{1}{T_v} \log \bigg[\frac{ \sinh (T_u l_u)}{2T_u T_v \rho_\infty^{-1}}\bigg]. 
  }
This expression for the bench agrees with the results obtained in~\cite{Apolo:2020qjm} where $\gamma$ was shown to correspond to the set of fixed points of the bulk modular flow generator. We also note that the bench of the swing surface always lies outside of the horizon of the BTZ black hole which is located at $\rho = 2 T_u T_v$.

The holographic entanglement entropy obtained from the area of the swing surface $\gamma_\mA = \gamma_- \cup \gamma \cup \gamma_+$ is then given by
   \eq{
    S_{\cal A} = \frac{\textrm{Area}(\gamma_{\cal A})}{4 G} = \frac{T_v l_v}{4G} +\frac{1}{4G} \log \bigg[ \frac{ \sinh (T_u l_u)} {2 T_uT_v \rho_\infty^{-1}}\bigg], \label{wcft:ee}
   }
which matches the entanglement entropy of WCFTs at the boundary~\eqref{wcft:eezero} provided that $\rho_\infty$ is related to the WCFT cutoff $\varepsilon$ by 
\eq{ 
\rho_\infty = \frac{2 T_v}{\varepsilon}. \label{wcft:cutoffs}
} 
The relationship~\eqref{wcft:cutoffs} between the radial cutoff in the bulk and the UV cutoff at the boundary can be derived, for example, from the Rindler transformation found in ref.~\cite{Song:2016gtd}. 

To conclude, we have explicitly checked that the holographic entropy proposal~\eqref{proposal} reproduces the entanglement entropy of single intervals in the AdS$_3$/WCFT correspondence. In particular, the swing surface $\gamma_\mA = \gamma_- \cup \gamma \cup \gamma_+$ described by the ropes~\eqref{wcft:wcftsols} and the bench~\eqref{wcft:gamma} reproduces the geometric picture previously obtained from the Rindler method in~\cite{Song:2016gtd}.


\subsection{Swing surfaces as RT/HRT surfaces in AdS$_3$/CFT$_2$} \label{se:sswads}

Let us now comment on swing surfaces in the AdS$_3$/CFT$_2$ correspondence. In order to construct a swing surface in AdS$_3$/CFT$_2$, we first need to find the set of null geodesics whose tangent vector reduces to the approximate modular flow generator of two-dimensional CFTs at the boundary. It is not difficult to check that null geodesics with $p_v=0$ have tangent vectors \eqref{wcft:xi2} that are not compatible with the approximate modular flow generator of CFTs, the latter of which are given in \eqref{wcft:approxmodflowcft}. Therefore, we must consider null geodesics with $p_u p_v > 0$, whose tangent vectors are given near the asymptotic boundary by (see Appendix~\ref{app:geodesicsads})
  \eq{
    \xi^{(\pm)} = \mp 2\pi \big[ ( u - u_\pm) \p_u + (v - v_\pm) \p_v - 2 \rho_\infty \p_\rho \big] + subleading. \label{wcft:pupvxi}
  }
The $u$ and $v$ components of the vectors \eqref{wcft:pupvxi} and \eqref{wcft:approxmodflowcft} differ by a relative sign, unless the null geodesics are pushed strictly to infinity, in which case these terms vanish exactly. Thus, in the AdS$_3$/CFT$_2$ correspondence the ropes of the swing surface can be interpreted as null geodesics that have shrunk all the way to the boundary. In particular, the extremal surface lying between the ropes is the RT/HRT surface and the swing surface proposal~\eqref{proposal} reduces to the standard RT/HRT prescription.


\subsection{Warped AdS$_3$} \label{se:wads}

The proposal for holographic entanglement entropy described in Section~\ref{se:generalproposal} can be generalized to warped AdS$_3$ spacetimes that are solutions to some massive gravity theories~\cite{Deser:1981wh,Deser:1982vy,Bergshoeff:2009hq} as well as Einstein gravity with additional matter fields~\cite{Detournay:2012dz}. The zero-mode warped AdS$_3$ backgrounds can be written in the same gauge used in \eqref{wcft:zeromode} such that
\eq{
ds^2 = \frac{(1+\alpha^2 T_v^2)d\rho^2}{4(\rho^2 - 4T_u^2 T_v^2)} + \rho\, du dv + \bigg[ T_u^2 - \frac{\alpha^2(\rho^2 - 4 T_u^2 T_v^2)}{4}  \bigg] du^2 + T_v^2 dv^2, \label{wcft:wadsmetric}
}
where $\alpha$ denotes the warping parameter. When $\alpha \to 0$ we recover the locally AdS$_3$ backgrounds described in eq.~\eqref{wcft:zeromode}.

Let us consider a single interval $\mA$ at the boundary whose endpoints are parametrized by~\eqref{wcft:intervalbulk}. In order to determine the holographic entanglement entropy we must first determine the ropes of the swing surface. The latter correspond to null geodesics whose tangent vectors satisfy the boundary condition $\xi^{(\pm)}|_{\p\cal M}=\zeta^{(\pm)}$. In this way, we find that the ropes are given by 
 	\renewcommand{\arraystretch}{.5}
\eq{
   \gamma_\pm:  \, \left\{ \begin{array}{l}
    \rho = \frac{4 |p_u| T_v \lambda}{1 + \alpha^2 T_v^2} + \rho_0 , \\
  \\
  u = \mp \frac{1}{4 T_u} \log\bigg(\dfrac{\rho + 2 T_u T_v}{\rho - 2 T_u T_v} \bigg) + {u}_\pm, \\
  \\
  v = \mp \frac{1}{4 T_v} \log\big[( \rho^2 - 4 T_u^2 T_v^2)/\rho^2_\infty \big]  + {v}_{\pm},
  \end{array}\right. \label{wcft:wcftsols2}
  }
which are the same null geodesics of the locally AdS$_3$ backgrounds given in \eqref{wcft:wcftsols} after a rescaling of the affine parameter $\lambda \to (1 + \alpha^2 T_v^2)\lambda$. It is not difficult to check that the vectors $\xi^{(\pm)}$ tangent to the ropes $\gamma_\pm$ are given by \eqref{wcft:xi2} and that their $u$ and $v$ components match those of the approximate modular flow generator of WCFTs at the boundary~\eqref{wcft:approxmodflowbulk}.

We now find the extremal surface lying between the $\gamma_+$ and $\gamma_-$ ropes of the swing surface. In general, finding the geodesic between two arbitrary points in the warped AdS$_3$ background~\eqref{wcft:wadsmetric}, or a general expression for the geodesic distance, is a complicated task. Nevertheless, we find that it is possible to determine the bench without using the explicit expression for the geodesic distance between two points. In warped AdS$_3$, the geodesic distance must be invariant under the action of its local isometry group, namely $SL(2,R)_L \times U(1)_R$, where the $U(1)_R$ factor corresponds to translations along $v$ while the $SL(2,R)_L$ generators are given in footnote~\ref{wcft:sl2}. In the previous section, we found that there are two independent functions $D_+(p_1, p_2)$ and $D_-(p_1, p_2)$ that are invariant under $SL(2,R)_L \times U(1)_R$ transformations. As a result, the geodesic distance $L_\alpha(p_1, p_2)$ depends on the points $p_1 $ and $p_2$ only through the functions $D_+(p_1, p_2)$ and $D_-(p_1, p_2)$ given in \eqref{wcft:Ddef} such that
\eq{
  L_\alpha (p_1, p_2) = L_\alpha (D_+, D_-).
  }
As a special case, for the locally AdS$_3$ backgrounds obtained by setting $\alpha=0$, we have $L_0(D_+ , D_-) = \cosh^{-1}(D_+ + D_-)$ as described in detail in Section~\ref{se:ssads}. We also note that the $D_\pm(p_1, p_2)$ functions are independent of the warping parameter $\alpha$. 

Let us take $p_1 \in \gamma_-$, $p_2 \in \gamma_+$, and use the parametrization of the null geodesics introduced in eqs.~\eqref{wcft:gammaplus} and~\eqref{wcft:gammaminus}. It is useful to note that the null geodesics $\gamma_\pm$ satisfy the following equation
\eq{
\p_{u_1} D_-(u_1, u_2) = \p_{u_2} D_-(u_1, u_2) = 0,
}
for both AdS$_3$ and warped AdS$_3$ spacetimes. This follows from the fact that $D_-(p_1, p_2)$ is a constant independent of the $u_1$ and $u_2$ parameters, cf.~eq.~\eqref{wcft:Dminus}. We thus find that the extremality conditions reduce to
   \eq{
   \p_{u_1} L_\alpha (u_1, u_2) = \p_{D_+}L_\alpha (D_+, D_-)\, \p_{u_1} D_+(u_1, u_2) = 0, \label{wcft:wadscondition1} \\
   \p_{u_2} L_\alpha (u_1, u_2) = \p_{D_+}L_\alpha (D_+, D_-)\, \p_{u_2} D_+(u_1, u_2)= 0. \label{wcft:wadscondition2}
   }
It follows that the solutions to $\p_{u_1} D_+(u_1, u_2) = \p_{u_2} D_+(u_1, u_2) = 0$ automatically solve the extremality conditions~\eqref{wcft:wadscondition1} and~\eqref{wcft:wadscondition2}. This means that the bench of the swing surface is the same in both AdS$_3$ as well as warped AdS$_3$ spacetimes, and is given by eq.~\eqref{wcft:gamma}. The area of the swing surface $\gamma_\mA = \gamma_- \cup \gamma \cup \gamma_+$ is then given by
   \eq{
    S_{\cal A} = \frac{\textrm{Area}(\gamma_{\cal A})}{4 G} = \frac{T_v l_v}{4G} +\frac{1}{4G} \log \bigg[ \frac{ \sinh (T_u l_u)} {2 T_uT_v \rho_\infty^{-1}}\bigg], \label{wcft:ee2}
   }
which agrees with the AdS$_3$ result~\eqref{wcft:ee} and matches the entanglement entropy of single intervals in WCFTs where the bulk and boundary cutoffs are related by~\eqref{wcft:cutoffs}. 

In eqs.~\eqref{wcft:wadscondition1} and \eqref{wcft:wadscondition2} we have assumed that $\p_{D_-} L_\alpha (D_+, D_-)$ does not diverge, which can be explicitly verified to be the case in the $\alpha \to 0$ limit. In addition, it is possible that other extremal surfaces exist in WAdS$_3$ satisfying eqs.~\eqref{wcft:wadscondition1} and~\eqref{wcft:wadscondition2} with $\p_{D_+}L_\alpha (D_+, D_-) =0$. We note that this does not happen in AdS$_3$ where $\p_{D_+}L_0 (D_+, D_-)= [\sinh L(u_1,u_2)]^{-1} > 0$ for all real values of $u_+$ and $u_-$, which means that  the extremality condition is equivalent to $\p_{u_1} D_+(u_1, u_2) = \p_{u_2} D_+(u_1, u_2) = 0$. 


\bigskip

\section*{Acknowledgments}
We are grateful to Pankaj Chaturvedi, Bartek Czech, Stephane Detournay, Daniel Harlow, Juan Maldacena, Prahar Mitra, Max Riegler, and Herman Verlinde for helpful discussions. LA and WS thank the Kavli Institute of Theoretical Physics for hospitality and for providing a stimulating environment during the program ``Gravitational Holography''. LA also thanks the Institute for Advanced Study for their kind hospitality. The work of LA, WS, and YZ was supported by the National Thousand-Young-Talents Program of China, NFSC Grant No.~11735001, and Beijing National Science Foundation No.~Z180003. LA and WS were supported in part by the National Science Foundation under Grant No.~NSF PHY-1748958. The work of LA was also supported by the International Postdoc Program at Tsinghua University and NFSC Grant No.~11950410499. HJ is supported by the Swiss National Science Foundation.

\appendix  

\section{Bulk modular flow and extremal surfaces}\label{app:extremal}

In this appendix we show that, in three spacetime dimensions, 
the set of  fixed points of a Killing vector $\xi$ extremizes the distance between two null geodesics $\gamma_\pm$ whose tangent vector is parallel to $\xi$.

As described in Section~\ref{se:rindler}, the bulk modular flow generator $\xi$ of the vacuum is an exact Killing vector that can be used to determine the swing surface $\gamma_\mA$ that is homologous to the interval $\mA$ at the boundary. In three spacetime dimensions, the swing surface is given by $\gamma_{\cal A} = \gamma_-\cup\gamma\cup\gamma_+$ where the ropes $\gamma_\pm$ are null geodesics emanating from the endpoints of $\mA$ that lie tangent to the modular flow. Let $q_\pm $ denote the intersection between these null geodesics and the set of fixed points $\gamma_\xi$  of the bulk modular flow generator. Then, the bench $\gamma$ of the swing surface is the subset of $\gamma_\xi$ that lies between the points $q_+$ and $q_-$.

The modular flow $e^{s\xi}$ maps any point $p \in \mM$ to a one-parameter family of points $p(s)$ where $p(0)=p$. Under the modular flow, points on $\gamma_\pm$ remain on $\gamma_\pm$ and the set of fixed points $\gamma_\xi$ are left invariant. In particular, we have
\eq{
 q_\pm(s) = q_\pm, \qquad \forall s \in \mathbb R. \quad 
  \label{app:fixedpoint}
}
Let us now consider the geodesic distance $L(p_1, p_2)$ between two points $p_1$ and $p_2$. Since $\xi$ is an exact Killing vector, the geodesic distance is invariant under the simultaneous action of $e^{s\xi}$ on the points $p_1$ and $p_2$, namely
\eq{
L\big(p_1(s),p_2(s)\big) = L (p_1, p_2 ). \label{app:disinv}
}
Consequently, we find that for any point $p_+ \in \gamma_+$ the geodesic distance $L(p_+, q_-)$ is independent of $p_{+}$ and is given by the distance between the two fixed points
 \eq{
L (p_+, q_-) = L\big(q_+,q_-\big), \quad \forall p_+ \in \gamma_+, \label{app:Dpq}  
}
where we have assumed that $L (p_+, q_-)$ is a continuous function of $p_+$. Similarly,  the distance between the fixed point $q_+$ and any point $p_-\in \gamma_-$ is independent of $p_-$. This guarantees that the points $q_+$ and $q_-$ extremize the distance $L(p_+, p_-)$ between any two points $p_+ \in \gamma_+$ and $p_- \in \gamma_-$. As a result, the bench of the swing surface extremizes the geodesic distance between the ropes $\gamma_+$ and $\gamma_-$.


\section{Parametrization of null geodesics}\label{app:nullgeodesics} 
In this appendix we show that it is always possible to parametrize a null geodesic in terms of a parameter $\tau$ such that the tangent vector  $\xi^\mu = 2\pi {d x^{\mu}(\tau)}/{d\tau}$ satisfies \eqref{s2:nmbk}, namely
\eq{
\quad \xi^{\mu}\nabla_{\mu}\xi^{\nu}=\pm  2\pi \xi^{\nu}.
}
By definition, $\tau$ is not an affine parameter; nevertheless, it can be related to an affine parameter $\lambda$ via the reparametrization $\lambda = f(\tau)$, where $x^{\mu}(\lambda)$ satisfies the geodesic equation
\eq{
	\frac{d x^{\mu}}{d\lambda} \nabla_{\mu} \Big( \frac{d x^{\nu}}{d\lambda} \Big)=0.
}
In order to accomplish this we need to solve the ODE
\eq{
\Big(\frac{d^2\tau}{d\lambda^2} \Big) \Big(\frac{d\tau}{d\lambda} \Big)^{-2} = \mp 1,
}
whose solution is given by
\eq{
\tau = \tau_0 \pm  \log| \lambda  -\lambda_0|. \label{app:odesol}
}
The null geodesic reaches the boundary when $\lambda\to\pm \infty$. For the upper plus sign in \eqref{app:odesol}, the boundary corresponds to $\tau \to +\infty$ and the tangent vector $\xi^\mu$ points towards the boundary; while for the lower minus sign, the boundary corresponds to $\tau \to -\infty$ and the tangent vector $\xi^\mu$ points towards the bulk. Using \eqref{app:odesol} we find that 
\eq{\label{xibdy}
\xi^\mu = 2\pi \frac{d x^{\mu}}{d\tau} = 2\pi \frac{d\lambda}{d\tau}\frac{d x^{\mu}}{d\lambda}=\pm 2\pi(\lambda - \lambda_0)\frac{d x^{\mu}}{d\lambda} ,
}
which shows that $\lambda=\lambda_0$ is a fixed point of $\xi$ provided that $dx^\mu/d\lambda$ does not diverge at $\lambda_0$.


\section{The approximate modular flow generator in the bulk} \label{app:aproxmodflowbulk}

In this appendix we show how the approximate bulk modular flow generator $\xi^{(p)}_\infty$ described in Section~\ref{se:localmodflow} can be determined in the models of non-AdS holography considered in this paper. 

\subsection{Flat$_3$/BMSFT} \label{app:aproxmodflowflat}

The approximate modular flow generator $\zeta^{(\pm)}$ of BMSFTs \eqref{localmodflow} can be extended into the bulk of the asymptotically flat spacetimes \eqref{flatbackground} by finding the linear combination of asymptotic Killing vectors that reduces to $\zeta^{(\pm)}$ near the endpoints of the boundary interval $\mA$. As discussed in Section~\ref{se:flat}, three-dimensional Einstein gravity with a vanishing cosmological constant admits an infinite number of asymptotic symmetries described by the three-dimensional BMS algebra. These symmetries are generated by the action of the asymptotic Killing vectors
\eq{
\eta = \big[ \epsilon(\phi) + u \p_\phi \sigma (\phi)\big] \partial_u + \sigma(\phi) \partial_\phi - r \p_\phi \sigma(\phi)  \p_r +{\cal O}(1/r), \label{app:flatask}
}
where $\epsilon(\phi)$ and $\sigma(\phi)$ are two arbitrary periodic functions. The approximate modular flow generator in the bulk is denoted by $\xi_\infty^{(\pm)}$ and can be obtained from the asymptotic Killing vectors~\eqref{app:flatask} by letting $\epsilon(\phi) = \pm 2\pi u_\pm$ and $\sigma(\phi) = \mp 2\pi ( \phi - \phi_\pm)$, whereupon
\eq{
\xi_\infty^{(\pm)} = \mp 2\pi \big[ (u - u_\pm) \p_u +  ( \phi - \phi_\pm) \p_\phi - r \p_r \big].
\label{app:amf}
}
By construction, the $u$ and $\phi$ components of~\eqref{app:amf} reproduce the approximate modular flow generator of BMSFTs at the boundary~\eqref{localmodflow}. We also note that the asymptotic Killing vectors~\eqref{app:amf} are tangent to the ropes of the swing surface, not only at the boundary, but also in the interior of the asymptotically flat spacetimes \eqref{flatbackground}. This follows from the fact that the $u$ and $v$ components of \eqref{app:amf} vanish along the ropes of the swing surface and that any vector proportional to $\p_r$ is null everywhere in the class of spacetimes \eqref{flatbackground}.


\subsection{AdS$_3$/WCFT} \label{app:aproxmodflowwarped}

We now consider the approximate modular flow generator in the bulk side of the AdS$_3$/WCFT correspondence. The asymptotic Killing vectors of AdS$_3$ gravity compatible with the CSS boundary conditions can be written as~\cite{Compere:2013bya}
  \eq{
  \eta = \epsilon(u)(\p_u + \p_v) + \frac{\sigma(u)}{2 T_v\sqrt{-c/ 6k}} \p_v  - \frac{r \epsilon'(u)}{2} \p_r + {\cal O}(1/r), \label{app:csskilling}
  }
where $\epsilon(u)$ and $\sigma(u)$ are two arbitrary functions. 
The approximate bulk modular flow generator $\xi^{(\pm)}_\infty$ can be obtained by finding the combination of $\epsilon(u)$ and $\sigma(u)$ functions for which the $u$ and $v$ components of \eqref{app:csskilling} match those of $\zeta^{(\pm)}$ in eq.~\eqref{wcft:approxmodflowbulk}. Using $\epsilon(u) = \mp 2\pi (u - u_\pm)$ and $\sigma(u) = - 2\pi\sqrt{-c/ 6k}$, we find that $\xi^{(\pm)}_\infty$ is given by
\eq{
\xi^{(\pm)}_\infty = \mp 2\pi (u -u_\pm ) (\partial_u +\partial_v) -\frac{\pi}{T_v} \partial_v \pm \pi r \partial_r + {\cal O}(1/r). \label{app:wcftbulkmodflow}
}
The norm of this asymptotic Killing vector is given by
\eqsp{
\! \xi^{(\pm)}_\infty  \cdot \xi^{(\pm)}_\infty & = \mp \,\,2\pi (u-u_\pm) \omega_\pm r^2  + \pi^2 + T_v^2 \omega_\pm^2 + 4\pi^2 (u- u_\pm)^2 L(u)  \\
& \hspace{12pt} \mp \frac{2\pi(u-u_\pm) \omega_\pm T_v^2 L(u)}{r^2} + \mathcal O(1/r^4), \label{app:wcftnorm}
}
where we used the general metric \eqref{wcft:adsmetric} and $\omega_\pm  \equiv \mp 2\pi(u-u_\pm)[1+ J'(u)] - \pi/T_v$.

From \eqref{app:wcftnorm} we find that the approximate modular flow generator $\xi^{(\pm)}_\infty$ is null, up to terms of $\mathcal O(1/r^4)$, on a surface $N_\pm$ that contains the endpoints of the boundary interval $\mA$ and is given by
 \eq{
N_\pm:  u =  u_\pm \mp \frac{T_v }{r^2} + \mathcal O(1/r^4). \label{app:lightsheet}
  }
  Given an endpoint $(u_\pm, v_\pm)$ at the boundary and the tangent vector $\xi^{(\pm)}_\infty$, we can solve for the null geodesic equations and find the ropes $\gamma_\pm$ of the swing surface for the more general backgrounds \eqref{wcft:adsmetric}. Once the ropes of the swing surface are known, we can extend the approximate modular flow generator $\xi^{\pm}_\infty$ into the interior of the spacetime. As a consistency check, we note that the null geodesics for the zero-mode backgrounds \eqref{wcft:zeromode} given in \eqref{wcft:wcftsols} lie on the light-sheet $N_\pm$ near the asymptotic boundary, and their tangent vectors \eqref{wcft:xi2} approach \eqref{app:wcftbulkmodflow}. Moreover, for these backgrounds, the light-sheets $N_\pm$ \eqref{app:lightsheet} have a closed-form expression, since the bulk modular flow generator is locally an exact Killing vector, and are given in the Fefferman-Graham gauge by \cite{Apolo:2020qjm}
\eq{
N_\pm:\, r^2 = \mp T_u T_v \coth\big[ T_u (u - u_\pm) \big].
}
%


\section{Parametrization of the swing surface in flat$_3$/BMSFT} \label{app:zeroMode2}

In this appendix we describe in more detail the parametrization of the swing surface in the flat$_3$/BMSFT correspondence.

\subsection{Comments on the swing surface for the Poincar\'{e} vacuum}\label{app:benchcommentsPoincare}

Let us begin by noting that a subtlety arises in the description of the bench of the swing surface for the Poincar\'e vacuum given in \eqref{flatX}. Intuitively, one might expect that $\zc\in [-{l_\zc/2},\,{l_\zc/2}]$ but this range does not accurately parametrize the bench since both the $u$ and $r$ coordinates in \eqref{flatX} diverge when $\zc \to 0$. In order to determine the actual range of $\zc$, it is more convenient to use Cartesian coordinates $(t,x,y)$ satisfying
\eq{
t &= (l_\zc^2 + 4\zc^2)  \frac{r}{4 l_\zc}+ \frac{2u}{l_\zc}, \qquad x =  \zc r+ \frac{l_u}{l_\zc}, \qquad y = (l_\zc^2 - 4\zc^2) \frac{r}{4 l_\zc} - \frac{2u}{l_\zc},  \label{tyxM0}
}
such that $ds^2=-dt^2 +dx^2+dy^2$. In these coordinates, the bench $\gamma$ is just a straight line parametrized by
\eq{
(t, x,y) = \Big( 0,0, -{l_u\over 2\zc}\Big) \qquad  \mathrm{where} \qquad   y_+ \le y \le y_-, \quad y_\pm= \mp {l_u\over l_\zc}. \label{bench}
}
When $y$ goes from $y_-$ to $0$ and then from $0$ to $y_+$, the $\zc$ coordinate goes from $-l_\zc/2$ to ${-}\infty $ and then continues from ${+}\infty$ to $l_\zc/2$. As a result, there seems to be a jump in $\zc$ from ${-}\infty$ to ${+}\infty$ as $y$ passes through zero. We can avoid this discontinuity by regularizing $y$ such that $y=-(l_u/2 R)  e^{-i\theta}$ where $R$ is large. It then follows that, as we go around the point $y = 0$, $\zc$ goes around a big circle at infinity $C_R: R e^{i\theta }$ with $\theta\in[0, \pi]$. The bench can then be parametrized by \eqref{bench} with the following range of $\zc$
  \eq{
  \zc \in [l_\zc/2,R) \cup C_R\cup  (-R, -l_\zc/2], \label{contour}
  }
which can be thought of as the complement of $\zc \in [-l_\zc/2, l_\zc/2]$, i.e.~the range of the interval $\mA$ at the boundary.

We learn two important facts from the discussion above: ($i$) a smooth bench cannot be parametrized in the range $\zc \in [-l_\zc/2, l_\zc/2]$ in Poincar\'e coordinates, but instead by the complementary range~\eqref{contour}; and ($ii$) integrals along the bench of the form $\int_{\gamma}d\zc f(\zc)$, where $f(\zc)$ is analytic in the upper-half complex plane, can be equivalently integrated over the range $\zc \in [-l_\zc/2, l_\zc/2]$. The second fact is useful in the calculation of the variation of the entanglement entropy. 


\subsection{The swing surface for zero-mode $M > 0$ backgrounds} \label{app:benchcomments}

In this appendix we describe the parametrization of the swing surface for the zero-mode asymptotically flat spacetimes considered in Section~\ref{se:zeromode}. 

\subsubsection*{The ropes}

\begin{figure}[ht]  
   \centering
    \includegraphics[scale=0.611]{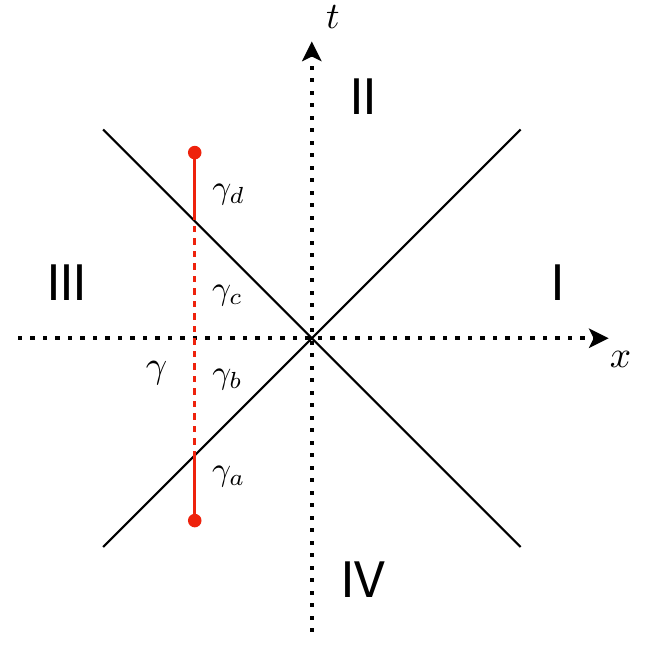}
    \caption{Four regions of a two-dimensional slice of Minkowski space. The Bondi coordinates with $r > 0$ cover only region II and part of region III. When $r \in (-\infty, \infty)$, the Bondi coordinates cover regions II, III, and IV. We also show the image of the bench under the map \eqref{transII}. For small $J$, the bench $\gamma=\gamma_a  \cup \gamma_b  \cup \gamma_c  \cup \gamma_d$  ends in Region II and IV.  The segment in the middle is dashed as it corresponds to analytically continued complex coordinates $(u,r,\phi)$.
 } \label{fig:patches}
 \end{figure}

As discussed in Section~\ref{se:flatlocalmodflow}, in a region sufficiently close to the boundary, the ropes $\gamma_\pm$ of the swing surface extend from the endpoints $\p\mA$ of $\mA$ along null geodesics tangent to $\p_r$. However, a subtlety arises when these geodesics extend into the interior of the zero-mode backgrounds \eqref{zeroMode} with $M > 0$  in the Bondi gauge. In order to see this, let us consider the following coordinate transformation between retarded Bondi coordinates $(u, \phi, r)$ and Cartesian coordinates $(t,x,y)$ such that $ds^2=-dt^2 +dx^2 +dy^2$,
\eq{
\begin{split}
 t&=\frac{1}{\sqrt M} \big[ r \cosh (\sqrt{M}\phi) -r_c \sinh(\sqrt{M}\phi) \big], \\
 x&=\frac{1}{\sqrt M}  \big[ r \sinh(\sqrt{M}\phi) -r_c \cosh (\sqrt{M}\phi)   \big], \\
 y&= \frac{1}{\sqrt{M}}\big( r-Mu -\sqrt{M} r_c \phi \big),
\end{split}\label{transII}
}
 where $r_c \equiv J/2\sqrt{M}$ is assumed to be non-negative. It is not difficult to see from the first two equations that the region $r^2>r_c^2$ is mapped to the region  $t^2-x^2>0$ in global Minkowski space and vice versa.  Furthermore, since $\phi$ satisfies
 \eq{
 \phi =  -\frac{1}{\sqrt M} \log \bigg[ \frac{\sqrt{M} (t-x)}{r+r_c} \bigg]= \frac{1}{\sqrt M} \log \bigg[ \frac{\sqrt{M} (t+x)}{r-r_c} \bigg], \label{phieq}
 }
requiring $\phi$ to be real implies that global Minkowski can be divided into four regions (see Fig.~\ref{fig:patches}):
\begin{itemize}
\item[] Region II:  \phantom{I}$r>r_c$, or equivalently $t-x>0$ and $t+x>0$,
\item[] Region III:  $-r_c<r<r_c$, or equivalently $t-x>0$ and $t+x<0$,
\item[] Region IV:  $\,r<-r_c$, or equivalently $t-x<0$ and $t+x<0$,
\item[] Region I:  \phantom{II} $t-x>0$ and $t+x<0$, not covered by \eqref{transII}.
\end{itemize}
Note that when $J=0$, region III can not be covered with real Bondi coordinates.  

In Cartesian coordinates, the null rope $\gamma_+$ can be written in region II as 
 	\renewcommand{\arraystretch}{.5}
\eq{
\gamma_+:\, \left\{ \begin{array}{l} 
t =\frac{1}{\sqrt M} \Big[ \lambda \cosh \Big({\sqrt{M}l_\phi\over2}\Big) -r_c \sinh\Big({\sqrt{M}l_\phi\over2}\Big) \Big], \\
\\
 x =\frac{1}{\sqrt M}  \Big[ \lambda \sinh\Big({\sqrt{M}l_\phi\over2}\Big) - r_c \cosh \Big({\sqrt{M}l_\phi\over2}\Big) \Big],  \\
 \\
 y= \frac{1}{\sqrt{M}} \Big( \lambda - M{l_u\over2} -\sqrt{M} r_c {l_\phi\over2} \Big),
 \end{array}\right. \label{cartesiancoords}
}
where $\lambda \ge 0$. The rope $\gamma_+$ can be extended to regions III and IV by letting $\lambda$ take negative values. Then, we can use the inverse coordinate transformation to express $\gamma_+$ in Bondi gauge. The inverse coordinate transformation is given by
\eq{
\begin{split}
 u &= \frac{1}{M} \big( r-\sqrt M y -\sqrt M r_c\phi \big) ,\\
  \phi &=  -\frac{1}{\sqrt M} \log \bigg[ \frac{\sqrt{M} (t-x)}{r+r_c} \bigg]= \frac{1}{\sqrt M} \log \bigg[ \frac{\sqrt{M} (t+x)}{r-r_c} \bigg], \\
  r &= \pm\sqrt{M(t^2-x^2)+r_c^2}, 
\end{split} \label{inversecartesian}
}
 where the minus sign is taken in regions III and IV for negative values of $r$. A similar analysis holds for the $\gamma_-$ rope such that the null geodesics $\gamma_\pm$ can be parametrized by
 \eq{
 \gamma_\pm:\, u=\pm{l_u\over2}, \quad \phi =  \pm{l_\phi\over2}, \quad r\in(-\infty,\infty). \label{flatropes}
 }

\subsubsection*{The bench}

Let us now describe the parametrization of the bench for the zero-mode backgrounds \eqref{zeroMode} with $M> 0$. In Cartesian coordinates, the two endpoints of $\gamma$ are located at $q_\pm=(t_\pm,\,x_\pm,\,y_\pm)$ where 
\eq{
t_\pm &= \mp{ \Big[\sqrt{M}\big(Jl_\phi +2Ml_u \big)\coth\big({\sqrt{M}l_\phi\over2}\big) -2J\Big]\over4M\sinh\big({\sqrt{M}l_\phi\over2}\big)}, \\
x_\pm &=-{ \big(Jl_\phi +2Ml_u\big)\over4\sqrt{M}\sinh({\sqrt{M}l_\phi\over2})},  \\
y_\pm &= \mp \frac{1}{4M} \Big[\sqrt{M} (J l_\phi + 2 M l_u) \coth^2\big( \tfrac{\sqrt{M} l_{\phi}}{2} \big) - 2 J \coth\big( \tfrac{\sqrt{M} l_{\phi}}{2} \big)\Big].
}
In Cartesian coordinates, the bench corresponds to a line that can be parametrized by
\eq{
\gamma:\quad x&= x_+, \qquad y ={y_+\over t_+} t, \qquad t\in[t_+,-t_+]. \label{benchcartesian}
}
In particular, the norm of a vector tangent to the bench is given by $\sinh(\sqrt{M}l_\phi/2)$ and is always positive so that the bench is spacelike. 

Using the inverse coordinate transformation~\eqref{inversecartesian}, we find that the bench \eqref{benchcartesian} in Bondi coordinates can be parametrized in terms of $\phi$ by
\eq{
  \!\!\!\! r(\phi) \!&= \Bigg[ \frac{{2J} \cosh (\sqrt{M} \phi) - \sqrt{M}(J l_{\phi} + 2 l_u M) \textrm{csch}\Big(\tfrac{\sqrt{M} l_{\phi}}{2} \Big)}{4\sqrt{M}\sinh(\sqrt{M}\phi)}  \Bigg],  \label{bondibenchr} \\
 \!\!\!\!\!\!  \!\!\!\! u(\phi) \!&= \!\frac{r(\phi)}{M} \!- \! \frac{J \phi}{2M} \!+\! \frac{\cosh\Big(\tfrac{\sqrt{M} l_\phi}{2}\Big)}{4M\sinh(\sqrt{M}\phi)} \bigg[ (J l_{\phi} \!+\! 2 l_u M) \textrm{csch} \Big(\tfrac{\sqrt{M} l_\phi}{2}\Big) \cosh (\sqrt{M}\phi)  \!-\! \frac{2J}{\sqrt{M}} \bigg].   \label{bondibenchu}
 }
Another subtlety arises with the range of the $\phi$ coordinate due to the fact that the bench \eqref{benchcartesian} spans different patches of Minkowski space as illustrated in Fig.~\ref{fig:patches}. For positive values of $l_u,\,l_\phi,\,M$ and $J$, we always have $t_++x_+<0$ and $t_--x_->0$. Furthermore, for sufficiently small $J$, one can see that $q_+$ is in region IV while $q_-$ is in region II. For large $J$, both $q_+$ and $q_-$ can be in region III, in which case $\gamma$ remains in region III. 

For simplicity, let us work out the range of $\phi$ in detail for the case where $J=0$. As we move along the bench from $q_+$ to $q_-$, $t$ increases from $t_+ < 0$ to zero and then to $t_- = |t_+|$.   Using the relations $r^2=M(t^2-x^2)$ and $\cosh (\sqrt {M}\phi)=\sqrt{M} {t / r}$ we find that in Bondi coordinates the bench $\gamma$ can be divided into four segments denoted by $\gamma_a$ through $\gamma_{d}$ such that (see Fig.~\ref{fig:patches}):
 
\begin{itemize}

\item in segment $\gamma_a$ with $t\in[t_+,x_+)$, the coordinate transformation~\eqref{transII} is well defined and increasing $t$ corresponds to increasing $r\in[-\sqrt{\smash[b]{M(t_+ ^2-x_+^2)}},0^-)$ and $\phi \in [ l_\phi/2,\infty)$;
\item in segment $\gamma_{b}$ with $t\in (x_+,0]$,
we have purely imaginary $r: i 0^-\to -i \sqrt{M}|x_+|$ and an imaginary part in $\phi: \infty -{i\pi\over2\sqrt{M}}\to 0^+  -{i\pi \over2\sqrt{M}}$;
\item in segment $\gamma_{ c}$ with $t\in[0,-x_+)$,  
we have purely imaginary $r: -i\sqrt{M}|x_+|\to -i0^+$ and an imaginary part in $\phi: 0^--{i\pi\over2\sqrt{M}} \to -\infty -{i\pi\over2\sqrt{M}}$;
\item in segment $\gamma_{d}$ with $t\in(-x_+,|t_+|]$, the coordinate transformation is real again, with $r \in (0^+, \sqrt{\smash[b]{M(t_- ^2-x_-^2)}}]$ and $\phi \in (-\infty, -{l_\phi/2}]$.
\end{itemize}

Note that we have to be careful when passing through $t = x_+$. When  $t\to x_++0^+$, we have $r\to 0^+$ and thus $\cosh (\sqrt {M}\phi)=+\infty$. On the other hand, when $t\to x_++0^-$, we have $r\to 0^-$ and $\cosh (\sqrt {M}\phi)=-\infty$, so $\phi $ must have an imaginary part $ \pm i \pi/\sqrt M $. The change in the imaginary part appears exactly at $r=0$ where the phase is ambiguous. In order to resolve the discontinuity in $\phi$, we can introduce a regularization by setting $r=\varepsilon e^{i \theta}$ for small enough $\varepsilon$. Then, when $r$ goes from $0^+$ to $0^-$, $\theta$ goes from $0$ to $\pi$. Consequently,  $\phi$  changes continuously from $\infty$ to $\infty -{i\pi\over2\sqrt{M}}\to 0^+  -{i\pi \over2\sqrt{M}}$. We will denote this part by ${\cal C}_+$. Similarly, there is also a segment ${\cal C}_-$ that connects $\gamma_{c}$ to $\gamma_{d}$. 

Finally, let us consider an integral along the bench  $\int_\gamma d\phi f(\phi)$ where $\gamma=\gamma_a \cup {\cal C}_+\cup \gamma_{b}\cup\gamma_{c} \cup {\cal C}_-\cup\gamma_{d}$. If the integrand $f(\phi)$ is an analytic function, then we can continuously deform this contour so that $\phi \in [-{l_\phi / 2}, {l_\phi/ 2}]$. Consequently, integrals along the bench can be equivalently evaluated along this range, namely
  \eq{
  \int_\gamma d\phi f(\phi) = \int_{-l_\phi/2}^{l_\phi/2} d\phi f(\phi).
  }

To summarize, the bench for zero-mode backgrounds with $M>0$ is described by   eqs.~\eqref{bondibenchr} and \eqref{bondibenchu}. Furthermore, for integrals on the bench, we can take the range of $\phi$ to be $[-{l_\phi/2},\,{l_\phi/2}]$.
 

\subsection{The swing surface for zero-mode $M < 0$ backgrounds} \label{app:swingzeromode2}

In this appendix we describe the parametrization of the swing surface for asymptotically flat zero-mode $M < 0$ backgrounds including the global Minkowski vacuum.

\subsubsection*{The ropes}

As discussed in Section~\ref{se:flatlocalmodflow}, the null geodesics $\gamma_{\pm}$ of the swing surface originate from the endpoints $\p\mA$ of the interval $\mA$ and are proportional to $\p_r$ near the asymptotic boundary. The description of these null geodesics as they move into the interior of the $M < 0$ backgrounds differs from that of $M> 0$ backgrounds, as they acquire $\phi$-dependence in the Bondi gauge. In order to see this it is convenient to introduce Cartesian coordinates satisfying $ds^2 = -dt^2 + dx^2 + dy^2$, 
   \eq{
   \begin{split}
   t &= \frac{1}{\sqrt{-M}} \big( r - M u - \sqrt{-M} r_c \phi\big), \\
   x &= \frac{1}{\sqrt{-M}} \big[ r \cos(\sqrt{-M}\phi) - r_c \sin(\sqrt{-M}\phi) \big], \\
    y &= \frac{1}{\sqrt{-M}} \big[ r \sin(\sqrt{-M}\phi) + r_c \cos(\sqrt{-M} \phi) \big],
    \end{split}
   }
where $r_c \equiv J/2\sqrt{-M}$ is real. It is not difficult to show that $x^2+y^2=(r^2+r_c^2)/(-M)$. As a result, when the angular momentum is finite, the Bondi coordinates cover only a part of Minkowski space where a circle of radius $r_c/\sqrt{-M}$ is excised from the $(x,y)$ plane in addition to a deficit angle. In Cartesian coordinates the null geodesics $\gamma_{\pm}$ correspond to straight lines originating at the asymptotic boundary at~\eqref{syminterval2} that are parametrized by the value of the radial coordinate such that
      	\renewcommand{\arraystretch}{.5}
\eq{
\gamma_\pm:\, \left\{ \begin{array}{l}
  t = \frac{1}{\sqrt{-M}} \Big( \lambda \mp M \frac{l_u}{2} \mp \sqrt{-M} r_c \frac{l_\phi}{2}\Big), \\
\\
   x = \frac{1}{\sqrt{-M}} \Big[ \lambda \cos\Big(\frac{\sqrt{-M} l_\phi}{2}\Big) \mp r_c \sin\Big(\frac{\sqrt{-M} l_\phi}{2}\Big) \Big], \\
   \\
    y = \frac{1}{\sqrt{-M}} \Big[ \pm \lambda \sin\Big(\frac{\sqrt{-M} l_\phi}{2}\Big) + r_c \cos\Big(\frac{\sqrt{-M} l_\phi}{2}\Big) \Big].
    \end{array} \right. \label{flatropescartesian}
   }
Note that the parameter $\lambda$ in eq.~\eqref{flatropescartesian} can take both positive and negative values and the asymptotic boundary is reached when $\lambda \to \infty$.

In Bondi gauge, the null geodesics $\gamma_{\pm}$ consist of two segments such that $ \gamma_\pm = \gamma_\pm^{(1)} \cup \gamma_\pm^{(2)}$. In the first segment the $u$ and $\phi$ coordinates are fixed and the radial coordinate varies from $r\to \infty$ to $r = 0$ such that 
 \eq{
 \gamma_{\pm}^{(1)}: \, u = \pm \frac{l_u}{2}, \quad \phi = \pm \frac{l_\phi}{2}, \quad r \in [0, \infty). \label{app:segment1}
 }
On the other hand, the $u$ and $\phi$ coordinates vary in the second segment of the ropes and the radial coordinate grows from $r = 0$ to $r\to\infty$. This segment can be parametrized by\footnote{When $J = 0$, the second segment is given by $\gamma_{\pm}^{(2)}: \, u = \pm\dfrac{l_u}{2} + \dfrac{r}{M}, \,\phi = \pm \dfrac{l_\phi}{2} + \dfrac{\pi}{\sqrt{-M}} , \, r \in [0, \infty)$. In this case, the value of the $\phi$ coordinate jumps discontinuously between the first and second segments. Hence, it is convenient to keep $J\ne 0$ as a regulator as it guarantees a smooth parametrization of the ropes. }

 	\renewcommand{\arraystretch}{.5}
\eq{
 \gamma_{\pm}^{(2)}: \left\{ \begin{array}{l}
  u =  \pm \dfrac{ l_{u}}{2} - \dfrac{J}{2 M} \Big(\phi \mp \dfrac{ l_{\phi}}{2}\Big) + \dfrac{r}{M}, \\
  \\
  r =  \dfrac{J}{\sqrt{-M}} \tan\bigg[ \dfrac{\sqrt{-M}(\phi  \mp \frac{ l_{\phi}}{2})}{2}\bigg],
  \end{array} \right. \qquad \phi \in  \Big[\pm\dfrac{l_\phi}{2}, \pm\dfrac{l_\phi}{2} + \dfrac{\pi}{\sqrt{-M}}\Big).  \label{app:segment2}
}
The two segments of the ropes are joined at the point $(u,\phi,r) = (\pm {l_{u}}/{2}, \pm {l_{\phi}/}{2}, 0 )$, which corresponds to the lower limit of the $\phi$ coordinate in \eqref{app:segment2}, namely $\phi = \pm l_\phi/2$. On the other hand, the upper limit of $\phi$ on the second segment $\gamma^{(2)}_\pm$ of the ropes $\phi \to \pm {l_{\phi}}/{2} + {\pi}/{\sqrt{-M}}$ corresponds to the point $(u,\phi,r) \to \big( \infty, \pm \frac{l_{\phi}}{2} + \frac{\pi}{\sqrt{-M}}, \infty\big)$.


\subsubsection*{The bench}

According to the holographic entropy proposal described in Section~\ref{se:proposal}, the bench $\gamma$ of the swing surface $\gamma_\mA$ is the spacelike geodesic that extremizes the distance between the $\gamma_+$ and $\gamma_-$ ropes. The bench obtained in this way can be parametrized in terms of the $\phi$ coordinate by
\eq{
r(\phi) &= -\frac{2 J\cos(\sqrt{-M}\phi) - \sqrt{-M}(J l_\phi + 2 M l_u) \csc \Big(\tfrac{\sqrt{-M} l_\phi}{2}\Big)   }{4\sqrt{-M} \sin(\sqrt{-M}\phi)  }, \label{bondibench1}  \\
 u(\phi) &= - \frac{J \phi}{2M} + \frac{r(\phi)}{M} - \frac{\cos\Big(\tfrac{\sqrt{-M} l_\phi}{2}\Big)}{4M\sin(\sqrt{-M}\phi)} \bigg[ (J l_{\phi} + 2 M l_u) \textrm{csc} \Big(\tfrac{\sqrt{-M} l_\phi}{2}\Big) \cos (\sqrt{-M}\phi) - \frac{2J}{\sqrt{-M}} \bigg], \label{bondibench2}
}
which correspond to the analytic continuation of eqs.~\eqref{bondibenchr} and~\eqref{bondibenchu}. The bench $\gamma$ intersects the first segment of $\gamma_+$ and the second segment of $\gamma_-$, or vice versa, depending on the value of the angular momentum with respect to the ``critical'' value $J_c$ defined by
  \eq{
  J_c = \frac{2   (-M)^{3/2}l_{u}}{\sqrt{-M} l_{\phi} - \sin(\sqrt{-M}  l_{\phi})}.
  }
When $J > J_c$, the bench ends on the first segment of $\gamma_+$ and the second segment of $\gamma_-$. Otherwise, when $J < J_c$, the bench ends on the second segment of $\gamma_+$ and the first segment of $\gamma_-$. Without loss of generality let us consider the $J < J_c$ case which includes the global Minkowski vacuum. In this case, we find that the range of $\phi$ parametrizing the location of the bench is given by
  \eq{
  \hphantom{-}\phi^*< \phi < \frac{2\pi}{\sqrt{-M}} - \frac{ l_{\phi}}{2},
    }
 where $\phi^* \ge  l_{\phi}/2$ and satisfies
  \eq{
  \phi^* = \frac{ l_{\phi}}{2} + \frac{2}{\sqrt{-M}} \tan^{-1}\Bigg\{ \frac{\csc \Big(\tfrac{ \sqrt{-M}l_{\phi}}{2}\Big)^2 \big[  J \sin(\sqrt{-M}  l_{\phi}) - \sqrt{-M}( J  l_{\phi} + 2 M  l_{u} )  \big]}{2J} \Bigg\}.
  }
  
Let us now consider the length of the swing surface $\gamma_\mA = \gamma_- \cup \gamma \cup \gamma_+$. Since the ropes $\gamma_{\pm}$ are null, only the bench $\gamma$ yields a nonvanishing contribution to the length of the swing surface. When $J < J_c$, the length $L$ of the bench is given by
  \eq{
  L = \int_{\gamma} \sqrt{g_{\mu\nu} dx^{\mu} dx^{\nu}} = \int_{\phi^*}^{2\pi/\sqrt{-M} -  l_{\phi}/2}  \sqrt{g_{\mu\nu} \p_\phi x^{\mu}  \p_\phi x^{\nu}}  d\phi.
  }
Despite the complicated value of $ \phi^*$ at which the bench and the rope intersect, we find that in both the $J< J_c$ and $J > J_c$ cases, the length of $\gamma$ is given by 
  \eq{
  L =\bigg| \sqrt{-M} \Big( l_u + \frac{J  l_{\phi}}{2M} \Big) \cot\Big(\frac{\sqrt{-M}  l_{\phi}}{2}\Big) - \frac{J}{M} \bigg|.
  }
Consequently, the holographic entanglement entropy reads
  \eq{
  S_{\mA} &= \frac{\textrm{Area}(\gamma_\mA)}{4G}  = \frac{1}{4G} \bigg| \sqrt{-M} \Big( l_u + \frac{J  l_{\phi}}{2M} \Big) \cot\Big(\frac{\sqrt{-M}  l_{\phi}}{2}\Big) - \frac{J}{M}\bigg|, \label{app:conicaldefectHEE}
  }
which corresponds to the analytic continuation of \eqref{flathee} and agrees with the independent derivations of entanglement entropy in refs.~\cite{Bagchi:2014iea,Jiang:2017ecm}. In particular, for the global Minkowski vacuum where $M=-1$ and $J = 0$, the holographic entanglement entropy is given by
\eq{
  S_{\mA}  =     \frac{l_u}{4G} \cot \Big( \frac{l_\phi}{2} \Big). \label{app:globalminkEE}
}


\section{Null geodesics in AdS$_3$} \label{app:geodesicsads}

In this appendix we describe the null geodesics of the zero-mode AdS$_3$ backgrounds~\eqref{wcft:zeromode} whose momenta satisfy $p_u p_v > 0$. 

Let us parametrize the null geodesics in terms of $(u,v,\rho)$ coordinates that depend on an affine parameter $\lambda$. The  geodesics satisfy the conservation equations~\eqref{wcft:geodeq1} reproduced here for convenience
  \eq{
 T_u^2 u' + {\rho v'\over2} = p_u, \qquad T_v^2 v' + {\rho u'\over2} = -p_v, \label{app:geodcond1}
  }
as well as the null constraint~\eqref{wcft:geodeq2} which is given by
\eq{
\frac{\rho'^2}{4(\rho^2 - 4T_u^2 T_v^2)} + T_u^2 u'^2 + T_v^2 v'^2 + \rho \,u' v' = 0. \label{app:geodcond2}
}
We assume the momenta $p_u$ and $p_v$ along the $u$ and $v$ coordinates satisfy $p_u p_v > 0$, which guarantees that the geodesics reach the asymptotic boundary at large $\lambda$. The solution to eqs.~\eqref{app:geodcond1} and~\eqref{app:geodcond2} is then given by
\eqsp{
\rho &= \frac{p_u^2 p_v^2 ( 2\lambda + \lambda_0)^2 - p_u^2 T_v^2 - p_v^2 T_u^2}{p_u p_v},\\
u &= - \frac{1}{4 T_u}  \log \bigg\{  \frac{ p_v^2 [T_u + p_u (2 \lambda + \lambda_0)]^2 - p_u^2 T_v^2 }{ p_v^2 [T_u - p_u (2 \lambda + \lambda_0)]^2 - p_u^2 T_v^2}  \bigg\} + u_{0}, \\
v &= + \frac{1}{4 T_v}  \log  \bigg\{  \frac{ p_u^2 [T_v + p_v (2 \lambda + \lambda_0)]^2 - p_v^2 T_u^2 }{ p_u^2 [T_v - p_v (2 \lambda + \lambda_0)]^2 - p_v^2 T_u^2}  \bigg\}  + v_{0}, 
\label{app:gensol}
}
where $\lambda_0$, $u_0$, and $v_0$ are integration constants and $\lambda$ is the affine parameter. In particular, at the cutoff surface $\rho = \rho_\infty$ near the asymptotic boundary $\p \cal M$, the null geodesics \eqref{app:gensol} reach the point 
\eq{
    (u,v,\rho) = \bigg( u_{0} - \sqrt{\frac{p_v}{p_u \rho_\infty}} +\mathcal O\bigg(\frac{1}{\rho_\infty}\bigg),\, v_{0} +  \sqrt{\frac{p_u}{p_v \rho_\infty}}  + \mathcal O \bigg(\frac{1}{\rho_\infty}\bigg), \, \rho_\infty \bigg). \label{app:bc}
  }
We see that the null geodesics~\eqref{app:gensol} emanate from a point $(u_0, v_0)$ at the boundary. Furthermore, we find that at the cutoff surface, the vector $\xi \equiv 2\pi \lambda (dx^{\mu}/d\lambda) \partial_\mu$ tangent to \eqref{app:gensol} is given by
  \eq{
    \xi =  -2\pi \big[ ( u - u_0) \p_u + (v - v_0) \p_v - 2 \rho_\infty \p_\rho \big] + subleading. \label{app:xi1}
  }

We now note that null geodesics with $p_u p_v > 0$ cannot describe the ropes $\gamma_\pm$ of the swing surface in (W)AdS$_3$/WCFT. This is because the $u$ and $v$ components of $\xi$ in \eqref{app:xi1} vanish at the endpoints, while the approximate modular flow generator of a WCFT is finite, cf.~eq.~\eqref{wcft:approxmodflowbulk}. Null geodesics with $p_u p_v > 0$ cannot describe ropes in AdS$_3$/CFT$_2$ either. In order to show this we first note that the relative sign between the $\p_u$ and $\p_v$ terms in \eqref{app:xi1} is independent of the values of the momenta $p_u$ and $p_v$, or the relative sign in the expansion of the null geodesics along the $u$ and $v$ coordinates in \eqref{app:bc}. This means that for any null geodesic whose momenta satisfy $p_u p_v > 0$, the tangent vector \eqref{app:xi1} reduces to the generator of dilations near the asymptotic boundary. In contrast, the modular flow generator in the AdS$_3$/CFT$_3$ correspondence generates boost near the endpoints of the boundary interval.


\bibliographystyle{JHEP} 
\bibliography{BMS_WCFT,refs}

\providecommand{\href}[2]{#2}\begingroup\raggedright\begin{thebibliography}{10}

\bibitem{Ryu:2006bv}
S.~Ryu and T.~Takayanagi, \emph{{Holographic derivation of entanglement entropy
  from AdS/CFT}},
  \href{https://doi.org/10.1103/PhysRevLett.96.181602}{\emph{Phys. Rev. Lett.}
  {\bfseries 96} (2006) 181602}
  [\href{https://arxiv.org/abs/hep-th/0603001}{{\ttfamily hep-th/0603001}}].

\bibitem{Hubeny:2007xt}
V.~E. Hubeny, M.~Rangamani and T.~Takayanagi, \emph{{A Covariant holographic
  entanglement entropy proposal}},
  \href{https://doi.org/10.1088/1126-6708/2007/07/062}{\emph{JHEP} {\bfseries
  07} (2007) 062} [\href{https://arxiv.org/abs/0705.0016}{{\ttfamily
  0705.0016}}].

\bibitem{Swingle:2009bg}
B.~Swingle, \emph{{Entanglement Renormalization and Holography}},
  \href{https://doi.org/10.1103/PhysRevD.86.065007}{\emph{Phys. Rev.}
  {\bfseries D86} (2012) 065007}
  [\href{https://arxiv.org/abs/0905.1317}{{\ttfamily 0905.1317}}].

\bibitem{VanRaamsdonk:2009ar}
M.~Van~Raamsdonk, \emph{{Comments on quantum gravity and entanglement}},
  \href{https://arxiv.org/abs/0907.2939}{{\ttfamily 0907.2939}}.

\bibitem{VanRaamsdonk:2010pw}
M.~Van~Raamsdonk, \emph{{Building up spacetime with quantum entanglement}},
  \href{https://doi.org/10.1007/s10714-010-1034-0,
  10.1142/S0218271810018529}{\emph{Gen. Rel. Grav.} {\bfseries 42} (2010) 2323}
  [\href{https://arxiv.org/abs/1005.3035}{{\ttfamily 1005.3035}}].

\bibitem{Casini:2011kv}
H.~Casini, M.~Huerta and R.~C. Myers, \emph{{Towards a derivation of
  holographic entanglement entropy}},
  \href{https://doi.org/10.1007/JHEP05(2011)036}{\emph{JHEP} {\bfseries 05}
  (2011) 036} [\href{https://arxiv.org/abs/1102.0440}{{\ttfamily 1102.0440}}].

\bibitem{Lewkowycz:2013nqa}
A.~Lewkowycz and J.~Maldacena, \emph{{Generalized gravitational entropy}},
  \href{https://doi.org/10.1007/JHEP08(2013)090}{\emph{JHEP} {\bfseries 08}
  (2013) 090} [\href{https://arxiv.org/abs/1304.4926}{{\ttfamily 1304.4926}}].

\bibitem{Dong:2016hjy}
X.~Dong, A.~Lewkowycz and M.~Rangamani, \emph{{Deriving covariant holographic
  entanglement}}, \href{https://doi.org/10.1007/JHEP11(2016)028}{\emph{JHEP}
  {\bfseries 11} (2016) 028}
  [\href{https://arxiv.org/abs/1607.07506}{{\ttfamily 1607.07506}}].

\bibitem{Anninos:2013nja}
D.~Anninos, J.~Samani and E.~Shaghoulian, \emph{{Warped Entanglement Entropy}},
  \href{https://doi.org/10.1007/JHEP02(2014)118}{\emph{JHEP} {\bfseries 02}
  (2014) 118} [\href{https://arxiv.org/abs/1309.2579}{{\ttfamily 1309.2579}}].

\bibitem{Castro:2015csg}
A.~Castro, D.~M. Hofman and N.~Iqbal, \emph{{Entanglement Entropy in Warped
  Conformal Field Theories}},
  \href{https://doi.org/10.1007/JHEP02(2016)033}{\emph{JHEP} {\bfseries 02}
  (2016) 033} [\href{https://arxiv.org/abs/1511.00707}{{\ttfamily
  1511.00707}}].

\bibitem{Song:2016pwx}
W.~Song, Q.~Wen and J.~Xu, \emph{{Generalized Gravitational Entropy for Warped
  Anti--de Sitter Space}},
  \href{https://doi.org/10.1103/PhysRevLett.117.011602}{\emph{Phys. Rev. Lett.}
  {\bfseries 117} (2016) 011602}
  [\href{https://arxiv.org/abs/1601.02634}{{\ttfamily 1601.02634}}].

\bibitem{Song:2016gtd}
W.~Song, Q.~Wen and J.~Xu, \emph{{Modifications to Holographic Entanglement
  Entropy in Warped CFT}},
  \href{https://doi.org/10.1007/JHEP02(2017)067}{\emph{JHEP} {\bfseries 02}
  (2017) 067} [\href{https://arxiv.org/abs/1610.00727}{{\ttfamily
  1610.00727}}].

\bibitem{Azeyanagi:2018har}
T.~Azeyanagi, S.~Detournay and M.~Riegler, \emph{{Warped Black Holes in
  Lower-Spin Gravity}},
  \href{https://doi.org/10.1103/PhysRevD.99.026013}{\emph{Phys. Rev.}
  {\bfseries D99} (2019) 026013}
  [\href{https://arxiv.org/abs/1801.07263}{{\ttfamily 1801.07263}}].

\bibitem{Wen:2018mev}
Q.~Wen, \emph{{Towards the generalized gravitational entropy for spacetimes
  with non-Lorentz invariant duals}},
  \href{https://doi.org/10.1007/JHEP01(2019)220}{\emph{JHEP} {\bfseries 01}
  (2019) 220} [\href{https://arxiv.org/abs/1810.11756}{{\ttfamily
  1810.11756}}].

\bibitem{Apolo:2018oqv}
L.~Apolo, S.~He, W.~Song, J.~Xu and J.~Zheng, \emph{{Entanglement and chaos in
  warped conformal field theories}},
  \href{https://doi.org/10.1007/JHEP04(2019)009}{\emph{JHEP} {\bfseries 04}
  (2019) 009} [\href{https://arxiv.org/abs/1812.10456}{{\ttfamily
  1812.10456}}].

\bibitem{Chen:2019xpb}
B.~Chen, P.-X. Hao and W.~Song, \emph{{R\`enyi mutual information in
  holographic warped CFTs}},
  \href{https://doi.org/10.1007/JHEP10(2019)037}{\emph{JHEP} {\bfseries 10}
  (2019) 037} [\href{https://arxiv.org/abs/1904.01876}{{\ttfamily
  1904.01876}}].

\bibitem{Gao:2019vcc}
B.~Gao and J.~Xu, \emph{{Holographic entanglement entropy in AdS$_3$$/$WCFT}},
  \href{https://arxiv.org/abs/1912.00562}{{\ttfamily 1912.00562}}.

\bibitem{Bagchi:2014iea}
A.~Bagchi, R.~Basu, D.~Grumiller and M.~Riegler, \emph{{Entanglement entropy in
  Galilean conformal field theories and flat holography}},
  \href{https://doi.org/10.1103/PhysRevLett.114.111602}{\emph{Phys. Rev. Lett.}
  {\bfseries 114} (2015) 111602}
  [\href{https://arxiv.org/abs/1410.4089}{{\ttfamily 1410.4089}}].

\bibitem{Basu:2015evh}
R.~Basu and M.~Riegler, \emph{{Wilson Lines and Holographic Entanglement
  Entropy in Galilean Conformal Field Theories}},
  \href{https://doi.org/10.1103/PhysRevD.93.045003}{\emph{Phys. Rev.}
  {\bfseries D93} (2016) 045003}
  [\href{https://arxiv.org/abs/1511.08662}{{\ttfamily 1511.08662}}].

\bibitem{Hosseini:2015uba}
S.~M. Hosseini and {\'A}.~V{\'e}liz-Osorio, \emph{{Gravitational anomalies,
  entanglement entropy, and flat-space holography}},
  \href{https://doi.org/10.1103/PhysRevD.93.046005}{\emph{Phys. Rev.}
  {\bfseries D93} (2016) 046005}
  [\href{https://arxiv.org/abs/1507.06625}{{\ttfamily 1507.06625}}].

\bibitem{Jiang:2017ecm}
H.~Jiang, W.~Song and Q.~Wen, \emph{{Entanglement Entropy in Flat Holography}},
  \href{https://doi.org/10.1007/JHEP07(2017)142}{\emph{JHEP} {\bfseries 07}
  (2017) 142} [\href{https://arxiv.org/abs/1706.07552}{{\ttfamily
  1706.07552}}].

\bibitem{Hijano:2017eii}
E.~Hijano and C.~Rabideau, \emph{{Holographic entanglement and Poincar{\'e}
  blocks in three-dimensional flat space}},
  \href{https://doi.org/10.1007/JHEP05(2018)068}{\emph{JHEP} {\bfseries 05}
  (2018) 068} [\href{https://arxiv.org/abs/1712.07131}{{\ttfamily
  1712.07131}}].

\bibitem{Godet:2019wje}
V.~Godet and C.~Marteau, \emph{{Gravitation in flat spacetime from
  entanglement}},  \href{https://arxiv.org/abs/1908.02044}{{\ttfamily
  1908.02044}}.

\bibitem{Fareghbal:2019czx}
R.~Fareghbal and M.~H. Shalamzari, \emph{{First Law of Entanglement Entropy in
  Flat-Space Holography}},  \href{https://arxiv.org/abs/1908.02560}{{\ttfamily
  1908.02560}}.

\bibitem{Gentle:2017ywk}
S.~A. Gentle and S.~Vandoren, \emph{{Lifshitz entanglement entropy from
  holographic cMERA}},
  \href{https://doi.org/10.1007/JHEP07(2018)013}{\emph{JHEP} {\bfseries 07}
  (2018) 013} [\href{https://arxiv.org/abs/1711.11509}{{\ttfamily
  1711.11509}}].

\bibitem{Gentle:2015cfp}
S.~A. Gentle and C.~Keeler, \emph{{On the reconstruction of Lifshitz
  spacetimes}}, \href{https://doi.org/10.1007/JHEP03(2016)195}{\emph{JHEP}
  {\bfseries 03} (2016) 195}
  [\href{https://arxiv.org/abs/1512.04538}{{\ttfamily 1512.04538}}].

\bibitem{Sanches:2016sxy}
F.~Sanches and S.~J. Weinberg, \emph{{Holographic entanglement entropy
  conjecture for general spacetimes}},
  \href{https://doi.org/10.1103/PhysRevD.94.084034}{\emph{Phys. Rev.}
  {\bfseries D94} (2016) 084034}
  [\href{https://arxiv.org/abs/1603.05250}{{\ttfamily 1603.05250}}].

\bibitem{Dong:2018cuv}
X.~Dong, E.~Silverstein and G.~Torroba, \emph{{De Sitter Holography and
  Entanglement Entropy}},
  \href{https://doi.org/10.1007/JHEP07(2018)050}{\emph{JHEP} {\bfseries 07}
  (2018) 050} [\href{https://arxiv.org/abs/1804.08623}{{\ttfamily
  1804.08623}}].

\bibitem{Lewkowycz:2019xse}
A.~Lewkowycz, J.~Liu, E.~Silverstein and G.~Torroba, \emph{{$T \bar T$ and EE,
  with implications for (A)dS subregion encodings}},
  \href{https://arxiv.org/abs/1909.13808}{{\ttfamily 1909.13808}}.

\bibitem{Bondi:1962px}
H.~Bondi, M.~G.~J. van~der Burg and A.~W.~K. Metzner, \emph{{Gravitational
  waves in general relativity. 7. Waves from axisymmetric isolated systems}},
  \href{https://doi.org/10.1098/rspa.1962.0161}{\emph{Proc. Roy. Soc. Lond.}
  {\bfseries A269} (1962) 21}.

\bibitem{Sachs:1962wk}
R.~K. Sachs, \emph{{Gravitational waves in general relativity. 8. Waves in
  asymptotically flat space-times}},
  \href{https://doi.org/10.1098/rspa.1962.0206}{\emph{Proc. Roy. Soc. Lond.}
  {\bfseries A270} (1962) 103}.

\bibitem{Sachs:1962zza}
R.~Sachs, \emph{{Asymptotic symmetries in gravitational theory}},
  \href{https://doi.org/10.1103/PhysRev.128.2851}{\emph{Phys. Rev.} {\bfseries
  128} (1962) 2851}.

\bibitem{Barnich:2006av}
G.~Barnich and G.~Compere, \emph{{Classical central extension for asymptotic
  symmetries at null infinity in three spacetime dimensions}},
  \href{https://doi.org/10.1088/0264-9381/24/5/F01,
  10.1088/0264-9381/24/11/C01}{\emph{Class. Quant. Grav.} {\bfseries 24} (2007)
  F15} [\href{https://arxiv.org/abs/gr-qc/0610130}{{\ttfamily gr-qc/0610130}}].

\bibitem{Barnich:2010eb}
G.~Barnich and C.~Troessaert, \emph{{Aspects of the BMS/CFT correspondence}},
  \href{https://doi.org/10.1007/JHEP05(2010)062}{\emph{JHEP} {\bfseries 05}
  (2010) 062} [\href{https://arxiv.org/abs/1001.1541}{{\ttfamily 1001.1541}}].

\bibitem{Bagchi:2010eg}
A.~Bagchi, \emph{{Correspondence between Asymptotically Flat Spacetimes and
  Nonrelativistic Conformal Field Theories}},
  \href{https://doi.org/10.1103/PhysRevLett.105.171601}{\emph{Phys. Rev. Lett.}
  {\bfseries 105} (2010) 171601}
  [\href{https://arxiv.org/abs/1006.3354}{{\ttfamily 1006.3354}}].

\bibitem{Bagchi:2012cy}
A.~Bagchi and R.~Fareghbal, \emph{{BMS/GCA Redux: Towards Flatspace Holography
  from Non-Relativistic Symmetries}},
  \href{https://doi.org/10.1007/JHEP10(2012)092}{\emph{JHEP} {\bfseries 10}
  (2012) 092} [\href{https://arxiv.org/abs/1203.5795}{{\ttfamily 1203.5795}}].

\bibitem{Hofman:2011zj}
D.~M. Hofman and A.~Strominger, \emph{{Chiral Scale and Conformal Invariance in
  2D Quantum Field Theory}},
  \href{https://doi.org/10.1103/PhysRevLett.107.161601}{\emph{Phys. Rev. Lett.}
  {\bfseries 107} (2011) 161601}
  [\href{https://arxiv.org/abs/1107.2917}{{\ttfamily 1107.2917}}].

\bibitem{Detournay:2012pc}
S.~Detournay, T.~Hartman and D.~M. Hofman, \emph{{Warped Conformal Field
  Theory}}, \href{https://doi.org/10.1103/PhysRevD.86.124018}{\emph{Phys. Rev.}
  {\bfseries D86} (2012) 124018}
  [\href{https://arxiv.org/abs/1210.0539}{{\ttfamily 1210.0539}}].

\bibitem{Compere:2013bya}
G.~Comp\`ere, W.~Song and A.~Strominger, \emph{{New Boundary Conditions for
  AdS3}}, \href{https://doi.org/10.1007/JHEP05(2013)152}{\emph{JHEP} {\bfseries
  05} (2013) 152} [\href{https://arxiv.org/abs/1303.2662}{{\ttfamily
  1303.2662}}].

\bibitem{Anninos:2008fx}
D.~Anninos, W.~Li, M.~Padi, W.~Song and A.~Strominger, \emph{{Warped AdS(3)
  Black Holes}},
  \href{https://doi.org/10.1088/1126-6708/2009/03/130}{\emph{JHEP} {\bfseries
  03} (2009) 130} [\href{https://arxiv.org/abs/0807.3040}{{\ttfamily
  0807.3040}}].

\bibitem{Compere:2009zj}
G.~Compere and S.~Detournay, \emph{{Boundary conditions for spacelike and
  timelike warped AdS$_3$ spaces in topologically massive gravity}},
  \href{https://doi.org/10.1088/1126-6708/2009/08/092}{\emph{JHEP} {\bfseries
  08} (2009) 092} [\href{https://arxiv.org/abs/0906.1243}{{\ttfamily
  0906.1243}}].

\bibitem{Apolo:2020qjm}
L.~Apolo, H.~Jiang, W.~Song and Y.~Zhong, \emph{{Modular Hamiltonians in flat
  holography and (W)AdS/WCFT}},
  \href{https://doi.org/10.1007/JHEP09(2020)033}{\emph{JHEP} {\bfseries 09}
  (2020) 033} [\href{https://arxiv.org/abs/2006.10741}{{\ttfamily
  2006.10741}}].

\bibitem{ElShowk:2011ag}
S.~El-Showk and K.~Papadodimas, \emph{{Emergent Spacetime and Holographic
  CFTs}}, \href{https://doi.org/10.1007/JHEP10(2012)106}{\emph{JHEP} {\bfseries
  10} (2012) 106} [\href{https://arxiv.org/abs/1101.4163}{{\ttfamily
  1101.4163}}].

\bibitem{Bisognano:1975ih}
J.~J. Bisognano and E.~H. Wichmann, \emph{{On the Duality Condition for a
  Hermitian Scalar Field}}, \href{https://doi.org/10.1063/1.522605}{\emph{J.
  Math. Phys.} {\bfseries 16} (1975) 985}.

\bibitem{Bisognano:1976za}
J.~J. Bisognano and E.~H. Wichmann, \emph{{On the Duality Condition for Quantum
  Fields}}, \href{https://doi.org/10.1063/1.522898}{\emph{J. Math. Phys.}
  {\bfseries 17} (1976) 303}.

\bibitem{Cardy:2016fqc}
J.~Cardy and E.~Tonni, \emph{{Entanglement hamiltonians in two-dimensional
  conformal field theory}},
  \href{https://doi.org/10.1088/1742-5468/2016/12/123103}{\emph{J. Stat. Mech.}
  {\bfseries 1612} (2016) 123103}
  [\href{https://arxiv.org/abs/1608.01283}{{\ttfamily 1608.01283}}].

\bibitem{Casini:2017roe}
H.~Casini, E.~Teste and G.~Torroba, \emph{{Modular Hamiltonians on the null
  plane and the Markov property of the vacuum state}},
  \href{https://doi.org/10.1088/1751-8121/aa7eaa}{\emph{J. Phys.} {\bfseries
  A50} (2017) 364001} [\href{https://arxiv.org/abs/1703.10656}{{\ttfamily
  1703.10656}}].

\bibitem{Czech:2019vih}
B.~Czech, J.~De~Boer, D.~Ge and L.~Lamprou, \emph{{A Modular Sewing Kit for
  Entanglement Wedges}},  \href{https://arxiv.org/abs/1903.04493}{{\ttfamily
  1903.04493}}.

\bibitem{Lashkari:2016idm}
N.~Lashkari, J.~Lin, H.~Ooguri, B.~Stoica and M.~Van~Raamsdonk,
  \emph{{Gravitational positive energy theorems from information
  inequalities}}, \href{https://doi.org/10.1093/ptep/ptw139}{\emph{PTEP}
  {\bfseries 2016} (2016) 12C109}
  [\href{https://arxiv.org/abs/1605.01075}{{\ttfamily 1605.01075}}].

\bibitem{Dong:2017xht}
X.~Dong and A.~Lewkowycz, \emph{{Entropy, Extremality, Euclidean Variations,
  and the Equations of Motion}},
  \href{https://doi.org/10.1007/JHEP01(2018)081}{\emph{JHEP} {\bfseries 01}
  (2018) 081} [\href{https://arxiv.org/abs/1705.08453}{{\ttfamily
  1705.08453}}].

\bibitem{Dong:2013qoa}
X.~Dong, \emph{{Holographic Entanglement Entropy for General Higher Derivative
  Gravity}}, \href{https://doi.org/10.1007/JHEP01(2014)044}{\emph{JHEP}
  {\bfseries 01} (2014) 044} [\href{https://arxiv.org/abs/1310.5713}{{\ttfamily
  1310.5713}}].

\bibitem{Wald:1993nt}
R.~M. Wald, \emph{{Black hole entropy is the Noether charge}},
  \href{https://doi.org/10.1103/PhysRevD.48.R3427}{\emph{Phys. Rev.} {\bfseries
  D48} (1993) R3427} [\href{https://arxiv.org/abs/gr-qc/9307038}{{\ttfamily
  gr-qc/9307038}}].

\bibitem{Iyer:1994ys}
V.~Iyer and R.~M. Wald, \emph{{Some properties of Noether charge and a proposal
  for dynamical black hole entropy}},
  \href{https://doi.org/10.1103/PhysRevD.50.846}{\emph{Phys. Rev.} {\bfseries
  D50} (1994) 846} [\href{https://arxiv.org/abs/gr-qc/9403028}{{\ttfamily
  gr-qc/9403028}}].

\bibitem{Barnich:2001jy}
G.~Barnich and F.~Brandt, \emph{{Covariant theory of asymptotic symmetries,
  conservation laws and central charges}},
  \href{https://doi.org/10.1016/S0550-3213(02)00251-1}{\emph{Nucl. Phys.}
  {\bfseries B633} (2002) 3}
  [\href{https://arxiv.org/abs/hep-th/0111246}{{\ttfamily hep-th/0111246}}].

\bibitem{Asadi:2018lzr}
M.~Asadi and R.~Fareghbal, \emph{{Holographic Calculation of BMSFT Mutual and
  3-partite Information}},
  \href{https://doi.org/10.1140/epjc/s10052-018-6098-0}{\emph{Eur. Phys. J.}
  {\bfseries C78} (2018) 620}
  [\href{https://arxiv.org/abs/1802.06618}{{\ttfamily 1802.06618}}].

\bibitem{Wall:2012uf}
A.~C. Wall, \emph{{Maximin Surfaces, and the Strong Subadditivity of the
  Covariant Holographic Entanglement Entropy}},
  \href{https://doi.org/10.1088/0264-9381/31/22/225007}{\emph{Class. Quant.
  Grav.} {\bfseries 31} (2014) 225007}
  [\href{https://arxiv.org/abs/1211.3494}{{\ttfamily 1211.3494}}].

\bibitem{Grumiller:2019xna}
D.~Grumiller, P.~Parekh and M.~Riegler, \emph{{Local quantum energy conditions
  in non-Lorentz-invariant quantum field theories}},
  \href{https://arxiv.org/abs/1907.06650}{{\ttfamily 1907.06650}}.

\bibitem{Detournay:2020vrd}
S.~Detournay, D.~Grumiller, M.~Riegler and Q.~Vandermiers,
  \emph{{Uniformization of Entanglement Entropy in Holographic Warped Conformal
  Field Theories}},  \href{https://arxiv.org/abs/2006.16167}{{\ttfamily
  2006.16167}}.

\bibitem{Barnich:2012aw}
G.~Barnich, A.~Gomberoff and H.~A. Gonzalez, \emph{{The Flat limit of three
  dimensional asymptotically anti-de Sitter spacetimes}},
  \href{https://doi.org/10.1103/PhysRevD.86.024020}{\emph{Phys. Rev.}
  {\bfseries D86} (2012) 024020}
  [\href{https://arxiv.org/abs/1204.3288}{{\ttfamily 1204.3288}}].

\bibitem{Brown:1986nw}
J.~D. Brown and M.~Henneaux, \emph{{Central Charges in the Canonical
  Realization of Asymptotic Symmetries: An Example from Three-Dimensional
  Gravity}}, \href{https://doi.org/10.1007/BF01211590}{\emph{Commun. Math.
  Phys.} {\bfseries 104} (1986) 207}.

\bibitem{Compere:2008us}
G.~Compere and D.~Marolf, \emph{{Setting the boundary free in AdS/CFT}},
  \href{https://doi.org/10.1088/0264-9381/25/19/195014}{\emph{Class. Quant.
  Grav.} {\bfseries 25} (2008) 195014}
  [\href{https://arxiv.org/abs/0805.1902}{{\ttfamily 0805.1902}}].

\bibitem{Troessaert:2013fma}
C.~Troessaert, \emph{{Enhanced asymptotic symmetry algebra of AdS$_{3}$}},
  \href{https://doi.org/10.1007/JHEP08(2013)044}{\emph{JHEP} {\bfseries 1308}
  (2013) 044} [\href{https://arxiv.org/abs/1303.3296}{{\ttfamily 1303.3296}}].

\bibitem{Avery:2013dja}
S.~G. Avery, R.~R. Poojary and N.~V. Suryanarayana, \emph{{An
  sl(2,$\mathbb{R}$) current algebra from $AdS_3$ gravity}},
  \href{https://doi.org/10.1007/JHEP01(2014)144}{\emph{JHEP} {\bfseries 1401}
  (2014) 144} [\href{https://arxiv.org/abs/1304.4252}{{\ttfamily 1304.4252}}].

\bibitem{Apolo:2014tua}
L.~Apolo and M.~Porrati, \emph{{Free boundary conditions and the
  AdS$_3$/CFT$_2$ correspondence}},
  \href{https://doi.org/10.1007/JHEP03(2014)116}{\emph{JHEP} {\bfseries 1403}
  (2014) 116} [\href{https://arxiv.org/abs/1401.1197}{{\ttfamily 1401.1197}}].

\bibitem{Deser:1981wh}
S.~Deser, R.~Jackiw and S.~Templeton, \emph{{Topologically Massive Gauge
  Theories}}, \href{https://doi.org/10.1006/aphy.2000.6013,
  10.1016/0003-4916(82)90164-6}{\emph{Annals Phys.} {\bfseries 140} (1982)
  372}.

\bibitem{Deser:1982vy}
S.~Deser, R.~Jackiw and S.~Templeton, \emph{{Three-Dimensional Massive Gauge
  Theories}}, \href{https://doi.org/10.1103/PhysRevLett.48.975}{\emph{Phys.
  Rev. Lett.} {\bfseries 48} (1982) 975}.

\bibitem{Bergshoeff:2009hq}
E.~A. Bergshoeff, O.~Hohm and P.~K. Townsend, \emph{{Massive Gravity in Three
  Dimensions}},
  \href{https://doi.org/10.1103/PhysRevLett.102.201301}{\emph{Phys. Rev. Lett.}
  {\bfseries 102} (2009) 201301}
  [\href{https://arxiv.org/abs/0901.1766}{{\ttfamily 0901.1766}}].

\bibitem{Detournay:2012dz}
S.~Detournay and M.~Guica, \emph{{Stringy Schr\"odinger truncations}},
  \href{https://doi.org/10.1007/JHEP08(2013)121}{\emph{JHEP} {\bfseries 08}
  (2013) 121} [\href{https://arxiv.org/abs/1212.6792}{{\ttfamily 1212.6792}}].

\end{thebibliography}\endgroup

\end{document}